# Deciphering the origins and growth of supermassive black holes

Yash Aggarwal ★

*Lamont-Doherty Earth Observatory of Columbia University, Emeritus research associate (retired), Palisades, NY 10965, USA*

## ABSTRACT

We present a well-tested, theoretically supported empirical relation that helps decipher the origins, growth, and properties of supermassive black holes (SMBHs). Based on theoretical considerations and analysis of mass ($M_{\rm BH}$) versus age ($t$) distribution of high-redshift (>5.6) SMBHs, we get $M_{\rm BH}$ = Ms exp [14.6($t$ – 100)/$t$(Myr)], which gives the SMBH's seed mass Ms, and its derivative gives the instantaneous mass-accretion rate. It yields seeds of ∼20–420 $\mathrm{M_\odot}$ (solar masses) for the recently discovered SMBHs GNz11, CEERS_1019, and UHZ1, and ∼3 × $10^4$ $\mathrm{M_\odot}$ for the largest (1.24 × $10^{10}$ $\mathrm{M_\odot}$) high-$z$ SMBH. It is applied to 132446 SMBHs at $z$ < 2.4. The resultant seeds are classified based on size and likely formation mechanism: ∼54 per cent are classified as light (<350 $\mathrm{M_\odot}$) deemed to be Pop III remnants; ∼40 per cent as intermediate (350–2 × $10^3$ $\mathrm{M_\odot}$) and ∼6 per cent as heavier seeds (2 × $10^3$–3 × $10^4$ $\mathrm{M_\odot}$), both of which formed possibly by mergers of Pop III remnants. The direct collapse black hole (DCBH) mechanism is not required but not excluded. Furthermore, the results show the following. The mass-accretion rate increases exponentially from the seed's inception ($z$ ∼ 30), reaches a broad plateau at $z$ ∼ 8.5–6 coincident with the epoch of reionization, and decreases monotonically thereafter. Sub-Eddington accretion is the norm, except during the first ∼150 Myr SMBHs either experienced super-Eddington accretion or the radiative efficiency was much < 0.1. The largest seed can potentially grow via luminous accretion to (6.6 ± 2.2) × $10^{10}$ $\mathrm{M_\odot}$, consistent with a theoretical limit of ∼5 × $10^{10}$ $\mathrm{M_\odot}$. The Eddington ratio decreases and radiative efficiency increases as $z$ decreases, consistent with recent findings.

## 1. INTRODUCTION

The origins and growth of supermassive black holes (SMBHs), now thought to exist in the centres of most galaxies, remain unresolved enigmas. It is generally accepted that the progenitor seeds of SMBHs formed or were assembled in the very early universe. However, it remains unclear how massive these seeds were. Were they predominantly light or heavy, or somewhere in between? The origins of black holes (BHs) or the mechanisms by which they may have formed are intimately related to seed mass. Proposed mechanisms by which seeds may have formed have been extensively reviewed, among others, by Latif & Ferrara (2016) and Volonteri, Habouzit & Colpi (2021) and can broadly be divided into three categories depending on seed mass. The three categories are light seeds with a typical mass of –100 $\mathrm{M_\odot}$ formed from the collapse of massive metal-free first stars (Madau & Rees 2001; Johnson & Bromm 2007) dubbed as Pop III remnants; heavy seeds $10^4$–$10^5$ $\mathrm{M_\odot}$ formed from the collapse of pristine gas clouds in massive dark matter haloes (Bromm & Loeb 2003; Begeleman et al. 2006; Lodato & Natarajan 2006; Shang et al. 2010) dubbed as DCBHs, or by hierarchical growth of BHs in dense stellar clusters (Davies et al. 2011); and intermediate-size (∼$10^3$ $\mathrm{M_\odot}$) seeds formed via runaway collisions of stars in dense stellar clusters (Portegies et al. 2004; Freitag et al. 2006; Mapelli, 2016) or by the hierarchical merger of BHs in stellar clusters (Davies et al. 2011; Lupi et al. 2014). Simulations attempting to understand the properties of the seeds formed by such mechanisms, drawbacks, and conditions necessary for their formation have also been extensively reviewed (see Latif & Ferrara 2016; Volonteri et al. 2021). More recently, however, Habouzit et al. (2022) performed cosmological hydrodynamical simulations to understand the co-evolution of BHs and their host galaxies; Bhowmick et al. (2024) studied BH seed formation in cosmological simulations using a new stochastic seed model; and Huang et al. (2020) examined the early growth of SMBHs in hydrodynamic simulations with different BH seeding scenarios. It is safe to conclude that there is no consensus as to which of the proposed mechanisms may have played a dominant role and presumably cannot be ascertained without knowing the mass distribution of the seeds formed.

Furthermore, BH seeds are thought to have formed at $z$ ∼ 25–30 (Barkana & Loeb, 2001), and the recent discovery of the AGN GNZ11 at $z$ = 10.6 (Maiolino et al. 2024) supports such an assumption. The existence of active galactic nuclei (AGN) exceeding $10^{10}$ $\mathrm{M_\odot}$ (solar masses) less than a billion years old (e.g. Wu et al. 2015) has, however, defied comprehension as to how these BHs grew to such massive sizes in such short times. One possible explanation is that SMBHs grew rapidly through galaxy mergers that supplied new gas reservoirs (e.g. Volonteri & Rees 2005; Chi-Hong Lin et al. 2023). Such a process is likely to be stochastic and unpredictable. The other possibility is through mergers of BHs. However, Pacucci &

★ E-mail: haggarwal@hotmail.com

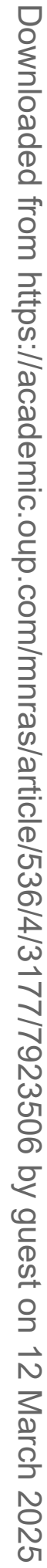

Loeb (2020), using observational data and theoretical models, studied the relative importance of gas accretion and mergers and concluded that mergers play a relatively minor role in the growth of SMBHs except at $z < 2$. Hence, the growth of SMBHs from progenitor seeds via luminous accretion remains the most likely viable path for the growth of SMBHs for much of their lifespan. The question is how these SMBHs, starting from progenitor seeds orders of magnitude smaller, grew to their enormous sizes in much less than a billion years. This dilemma is best illustrated by the following case studies of three recently discovered SMBHs at $z \geq 8.7$. Larson et al. (2023) concluded that CEERS_1019 at $z \sim 8.7$ and $M_{\rm BH} \sim 9 \times 10^6\,{\rm M}_\odot$ requires super-Eddington accretion from stellar-sized seeds or Eddington limited from massive BHs seeds. Schneider et al. (2023) and Maiolino et al. (2024) concluded that GNz11 at $z = 10.6$ and $M_{\rm BH} \sim 1.5 \times 10^6\,{\rm M}_\odot$ is accreting respectively at Eddington ratio $\lambda$ of $\sim 2.5$ and 5.5. In contrast, using simulations, Bhatt et al. (2024) found that the probability of observing a BH at $z \sim 10-11$ accreting with $\lambda \sim 5.5$ in the volume surveyed by *JWST* is $< 0.2$ per cent. And for the third AGN UHZ1 at $z \sim 10.1$ with an estimated $M_{\rm BH} \sim 4 \times 10^7\,{\rm M}_\odot$, Natarajan et al. (2024) concluded that it grew from a heavy seed ($\sim 10^4\,{\rm M}_\odot$) formed at $z \sim 14$ that was probably a DCBH. Again, it is safe to conclude that there is no consensus or clear understanding of how SMBHs, starting from seeds orders of magnitude smaller, grew to their observed sizes in several hundred million to less than a billion years.

While studies of individual AGNs provide valuable insights into the properties of that BH and may support a particular theory of the formation or growth of BH seeds, they cannot rule out other theories or channels of seed formation (Volonteri et al. 2021); nor do such studies provide measures of the growth rates of SMBHs. Presumably, solving the dual interrelated enigma of the origins and growth of SMBHs requires insights and constraints derived from large sets of observational data. The *JWST* may provide such data in years to come. Meanwhile, a large corpus of publicly available data, comprising known properties (redshift, mass $M_{\rm BH}$, Eddington ratio $\lambda$) of over a hundred thousand AGNs at various redshifts, remains largely unexploited. More precisely, this corpus consists of about 93 high-$z$ AGNs at $z > 5.6$ whose masses are well constrained and range over four orders of magnitude and $\sim$560 Myr in age, and 132 446 lower $z$ AGNs at $z < 2.4$ and $M_{\rm BH} \geq 10^7\,{\rm M}_\odot$ determined using Mg II lines by Kozlowski (2017). Fragione & Pacucci (2023) have attempted to constrain the distribution of BH seeds using a catalogue of 113 high-redshift ($z > 6$) AGNs compiled by Fan et al. (2023). Applying Bayesian analysis to these AGNs, they concluded that light and heavy seeds are required, whose distribution can be modelled by combining a power law and a lognormal function. Furthermore, they obtained mean values for the Eddington ratio, duty cycle, and radiative efficiency of the high-$z$ AGNs. Here, prompted by new insights garnered from a deconstruction of a commonly used theoretical formula for the growth of a BH, we analyse the mass versus age distribution of the 93 high-redshift AGNs and derive an empirical relation expressing a BH's mass $M_{\rm BH}$ as a function of its age $t$ or redshift $z$ and the mass Ms of its seed, from which the seed mass Ms can be ascertained knowing the BH's mass $M_{\rm BH}$ and redshift. The derivative of this primary relation gives a BH's accretion rate at any instant of its active life. Furthermore, using a BH's instantaneous accretion rate thus derived, its radiative efficiency can be obtained from its bolometric luminosity. Together, these relations comprise a set of powerful tools to decipher the origins, growth, and properties of SMBHs.

We begin by providing the theoretical basis of the primary relation and then derive the primary empirical relation using observational data. It is first applied to the 93 high-$z$ BHs. The resulting distributions of seed mass versus BH mass and BH age are analysed, and conclusions are drawn regarding the range of seeds necessary to account for the masses of the high-$z$ SMBHs observed thus far. In particular, the seed mass required to account for the mass of each of three recently discovered ultra-high $z$ AGNs, namely GNz11, CEERS-19, and UHZ1, is determined. The primary relation is then applied to the 132 446 AGNs at $z < 2.4$ to test its applicability and the validity of the conclusions drawn from its application to the high-$z$ AGNs. More importantly, the resulting mass distribution of the 132 446 seeds is analysed and compared to simulations of mass functions of the first massive stars by Hirano et al. (2014) and intermediate-size BHs formed via runaway collisions in nuclear star clusters by Devecchi et al. (2012) and gas-induced hierarchical growth of BHs by Lupi et al. (2014). Based on this comparison, the population of 132 446 seeds is classified into light, intermediate, and heavier, with corresponding mechanisms likely responsible for their formation. After that, the derivate of the primary relation is used to garner insights into a BH's accretion rate, the Eddington ratio, and the radiative efficiency as functions of redshift and find the role if any super-Eddington accretion plays in the growth of SMBHs. We end with a summary of the principal conclusions.

## 2. THEORETICAL BASIS

Conventionally, a BH's growth is defined by equation (1), where $M_{\rm BH}$ is the BH mass, $t$ its age (Myr), Ms its seed's mass, ts is the inception time, 45 Myr is the Salpeter time-scale, and $\gamma$ is a non-dimensional parameter defined in equation (2) and dubbed the 'growth efficiency factor' by Zubovas & King (2021).

$$M_{\rm BH}\,(\mathrm{t}) = \mathrm{Ms}\,\exp[\gamma\,(t - \mathrm{ts})\,/45\,(\mathrm{Myr})], \tag{1}$$

$$\gamma = [0.11\delta\lambda\,(1 - \varepsilon)\,/\varepsilon)]\,. \tag{2}$$

In equation (2), $\lambda$ is the Eddington ratio, $\varepsilon$ is the radiative efficiency, and $\delta$ is the duty cycle, all averaged over a BH's active life span ($t$ – ts). Note that for $\delta = 1$, $\varepsilon = 0.1$, and $\lambda = 1$, $\gamma = 1$ and equation (1) reduces to the conventional Salpeter relation for a BH accreting at the Eddington limit. By definition, $\lambda = L_{\rm bol}/L_{\rm EDD}$, and the bolometric luminosity $L_{\rm bol} \propto (\dot{M}\,\varepsilon/(1-\varepsilon)$ where $\dot{M}$ is the accretion rate, and the Eddington luminosity $L_{\rm EDD} \propto M_{\rm BH}$. Hence, $\gamma \propto \delta(\dot{M}/M_{\rm BH})$ or the accretion rate per unit BH mass The accretion rate per unit BH mass or $\gamma$ cannot be a constant because a BH of say $10^{10}\,{\rm M}_\odot$ at say $t = 10^9$ yr would grow untenably by $\sim$100 orders of magnitude by $t = 10^{10}$ yr. On the contrary, evidence shows that $\delta$ decreases as $z$ decreases (Shankar et al. 2010), and more recently Aggarwal (2024) has demonstrated that $\lambda$ decreases as $z$ decreases, but the radiative efficiency $\varepsilon$ increases as $z$ decreases, and hence $(1-\varepsilon)/\varepsilon)$ decreases as BH ages. Therefore, $\gamma$ or the accretion rate per unit BH mass must be an inverse function of a BH's age $t$, and its value averaged over the lifespan would invariably be greater than its value at any instant in the life of a BH. And since $\gamma$ is non-dimensional, we can simply define $\gamma = \beta 45/t$, where $\beta$ is a non-dimensional proportionality constant and $t$ in Myr, and 45 Myr is the standard e-folding time. Thus, equation (1) reduces to equation (3), where $M_{\rm BH}$ is a function of its known age $t$ and three unknowns Ms, $\beta$, and ts.

$$M_{\rm BH}\,(\mathrm{t}) = \mathrm{Ms}\,\exp[\beta\,(t - \mathrm{ts})\,/t]. \tag{3}$$

A SMBH's seed mass Ms can be ascertained using equation (3) if the values of the constant $\beta$ and the origin time ts of the seed were known. SMBH seeds are often assumed to have formed at $z \sim$ 25–30 when the universe was $\sim$100–130 Myr old (e.g. Barkana & Loeb 2001; Pacucci & Loeb 2020; Zubovas & King 2021; Fragione & Pacucci 2023). In particular, in their investigation of the growth of SMBHs, Pacucci & Loeb (2020) assert that the seeds formed at $z \sim 30$. Thus, assuming ts = 100 Myr, we can solve equation (3) and determine the value of $\beta$ if we find two SMBHs of different ages but with similar or identical seed masses. In attempting to ascertain the value of $\beta$, we shall adopt the assumption that SMBH seeds formed almost concurrently, and show a posteriori that it is justified. We do not, however, make any assumptions as to when the seeds formed and whether the BHs accreted at, above, or below the Eddington limit. Note that the value of ɣ determines whether a BH is accreting at, above, or below the Eddington limit, and it is not a constant but a function of $t$. In the following sections, we analyse the mass versus age distribution of 93 high-$z$ (>5.6) SMBHs and identify ~60 that are deemed to have similar seed masses. We use different subsets of this group of 60 SMBHs to obtain the value of $\beta$ and constrain the value of $t$s.

## 3. OBSERVATIONAL DATABASE

We searched the literature for SMBHs younger than a billion years ($z > 5.65$) whose masses are reliably known. Table A1 (Appendix) lists 59 SMBHs discovered until the end of 2022 with references for data sources. Where a BH had two or more estimates of $M_{\rm BH}$ from different sources, we generally chose the more recent determination. The average reported $1\sigma$ uncertainty in the $M_{\rm BH}$ of the 59 BHs is ~ 0.11 dex. In addition, Shen et al. (2019) list 50 SMBHs within a narrow $z$ range straddling $z = 6$, of which the $M_{\rm BH}$ of 12 are not well constrained, and 6 are duplicates of those in Table A1. The remaining 32 BHs extracted from table 3 in Shen et al. (2019) are identified in the Appendix. The $M_{\rm BH}$ of the large majority of the 91 BHs are based on the more reliable Mg II lines; only a few were obtained using the C v II emission lines. Shen et al. (2019), however, note that there may be a systematic error of a factor of up to 0.4 dex in $M_{\rm BH}$. This compilation of 91 BHs is comparable to that of Fan et al. (2023) for 113 BHs at $z > 5.3$, except that we indicate the uncertainty in $M_{\rm BH}$ of each BH and have excluded from the list those BHs whose $M_{\rm BH}$ are poorly constrained. In addition, there are two more recently discovered BHs at $z > 8.6$. These are GNz11 at $z = 10.6$ with $M_{\rm BH} \sim 1.5 \times 10^6\,{\rm M}_\odot$ and CEERS_1019 at $z \sim 8.7$ with $M_{\rm BH} \sim 9 \times 10^6\,{\rm M}_\odot$ for a total of 93. The UHZ1 AGN is omitted because its mass is estimated assuming $\lambda = 1$ (Natarajan et al. 2024) and hence uncertain. At lower redshifts, Kozlowski (2017) lists ~280 000 AGN at $z < 2.4$, almost all of which have $M_{\rm BH} \geq 10^7\,{\rm M}_\odot$. Of these 280 000 AGN, ~132 446 have $M_{\rm BH}$ determined using the more reliable Mg II lines and Eddington ratios based on a weighted average of bolometric luminosities derived using two or more AGN luminosities. The high-$z$ sample of 93 AGN appears to be skewed in favour of the larger BHs since ~75 per cent have $M_{\rm BH} \geq 10^9\,{\rm M}_\odot$, whereas the lower $z$ sample appears not to be because it has ~63 per cent in the $10^8$–$10^9\,{\rm M}_\odot$ range, ~27 per cent $\geq 10^9\,{\rm M}_\odot$, and the rest ~10 per cent < $10^8\,{\rm M}_\odot$. We note that there are probably a few hundred SMBHs at intermediate redshifts of $2.5 < z < 5.5$ that we did not compile because their inclusion would not alter the results of this study. Note that a BH's age $t$ corresponds to the cosmic time at the redshift $z$ at which it is observed and converted using the Hubble constant ($H_\odot = 67.4$ km s$^{-1}$ Mpc$^{-1}$) and matter density parameter ($\Omega_{\rm m} = 0.315$) from the Planck Collaboration VI (2020). An SMBH is defined as $\geq 10^6\,{\rm M}_\odot$. Throughout this paper, $M_{\rm BH}$ and Ms are in solar masses (${\rm M}_\odot$) and age $t$ in Myr.

## 4. EMPIRICAL RELATIONS

Fig. 1 shows the mass $M_{\rm BH}$ versus age $t$ distribution of 91 of the 93 SMBHs with density contours. The two recently discovered GNz11 and CEERS_19 at $t < 600$ Myr fall outside the bounds of Fig. 1 and are not plotted. The BHs in Fig. 1 are denoted by three symbols depending on BH mass to isolate those BHs with potentially similar mass seeds. Those in squares have similar masses within a factor of ~2 of $2.5 \times 10^9\,{\rm M}_\odot$ but different ages covering the entire age spectrum in Fig. 1; those in triangles nominally have $M_{\rm BH} > 5 \times 10^9\,{\rm M}_\odot$, and those in circles have $M_{\rm BH} < 10^9\,{\rm M}_\odot$. The two AGNs, GNz11, and CEERS-19, belong to the circles group and if plotted would fall to the left and below the left corner of Fig. 1 far from the rest of the BHs. The primary focus of Fig. 1 is on the group of BHs with similar masses but different ages (denoted by squares).

First, we note that ~13 BHs belonging to different groups with similar ages (in the middle of Fig. 1) straddling $t = 860 \pm 20$ Myr have markedly different $M_{\rm BH}$ spanning ~3 orders of magnitude. For each of these 13 BHs, we can write an equation (3) in which the term (ts – $t$) is nearly a constant assuming their seeds originated at nearly the same epoch. Hence, the exponent in equation (3) for each of these 13 BHs is nearly a constant irrespective of the value of $\beta$, and therefore their seed masses Ms differ from each other by the same proportion as their BH masses $M_{\rm BH}$. Notably, this group includes the largest (~$1.24 \times 10^{10}\,{\rm M}_\odot$; #31, Table A1) and the smallest (~$3.8 \times 10^7\,{\rm M}_\odot$; #27, Table A1) BH in Fig. 1. Their masses differ by ~3 orders of magnitude, and so must their seed masses Ms. It is likely that BH #31had a heavy seed of $10^4$–$10^5\,{\rm M}_\odot$, not only because it is the largest high-$z$ BH, but also because if it had a seed say an order of magnitude smaller it would imply that the smaller BH (#27) would have had a seed of ~ a solar mass which is unlikely. Thus, assuming that the seed of BH#31 had a mass of $10^4$–$10^5\,{\rm M}_\odot$, we can determine the corresponding values of $\beta$ for different values of seed origin time ts using equation (3). We get $\beta$ =13.25 and 15.85 for ts = 100 Myr ($z \sim 30$); $\beta$ = 15.22 and 18.21 for ts = 200 Myr ($z \sim 18$); and $\beta$ = 17.9 and 21.42 for ts = 300 Myr ($z \sim 14$). For these values of $\beta$, equation (3) predicts that a seed of $10^4$–$10^5\,{\rm M}_\odot$ originating at three different times, ts = 100, 200, and 300 Myr, would have grown for example at $z = 2$ to ~3.8–$4.3 \times 10^{10}\,{\rm M}_\odot$, ~1.62–$2.68 \times 10^{11}\,{\rm M}_\odot$, and ~1.15–$2.8 \times 10^{12}\,{\rm M}_\odot$, respectively. The predicted mass of the seed originating at ts = 100 Myr is comparable to the mass of the largest SMBHs at a similar redshift (TON 618 at $z = 2.219$ and $4.07 \times 10^{10}\,{\rm M}_\odot$; Xue Ge et al. 2019), whereas the other two predicted masses substantially exceed the masses of the largest SMBHs observed to date. The implication is that the likely value of ts is close to ~100 Myr rather than 200–300 Myr and that $\beta$ is likely to be between 13.25 and 15.85.

Fig. 1 also reveals that ~60 BHs (squares) have masses within a factor of 2 of $2.5 \times 10^9\,{\rm M}_\odot$ but different ages that range from ~ $t = 676$ to 1000 Myr. For any likely value of $\beta$, equation (3) indicates that within this age range the maximum difference in the masses of BHs of different ages but identical seed masses is at the most a factor of ~2. And since these BHs have similar masses, the implication is that these 60 BHs had similar-mass seeds. Hence, for each of the 60 BHs, we can write equation (3) with one variable ($t$) and three free parameters (Ms, ts, and $\beta$). We used the 'SANN' method (Belisle 1992), a Monte Carlo technique of solving optimization problems, with 10 million iterations to simultaneously solve the equations and optimize parameter values. Several sets of values for the parameters were generated using all 60 equations and subsets, allowing all three parameters to be free, assuming ts to be 100 150 and 200 Myr, and varying the time window within which the BHs had similar masses, the motivation being to obtain the most likely optimum value for $\beta$. Computations with ts = 150 and 200 gave the largest RMS residuals and were rejected. A covariance was noted between ts and $\beta$, in that an earlier ts gave a somewhat lower $\beta$ and vice versa. As suspected, subsets comprising BHs with masses within a narrower range of the mean value of 2.5 × 10$^9$ M$_\odot$ or a narrower age range produced lower residuals. The ensemble of the solutions indicated that the most likely value for ts was ~ 100 ± 30 Myr. Finally, ts = 100 Myr ($z$ ~ 30) and the corresponding $\beta$ ~14.6 were adopted as the most likely optimum values. It is noteworthy that these optimum values of $\beta$ and ts are within the bounds established above using the data for the largest SMBH not belonging to the group of 60. Furthermore, the preceding analyses indicate that if BH seeds had formed at substantially different epochs rather than almost concurrently as assumed, different BHs would have different $\beta$ and the 60 equations or subsets could not have converged to give a single $\beta$ value. Lastly, we estimate that if the BH masses are systematically underestimated by say a factor of 2, then the corresponding $\beta$ value would be ~14.8 with an origin time ts ~70 Myr, but the seed mass Ms for that BH would remain unchanged. Conversely, if the BH masses are systematically overestimated by a similar factor, the corresponding $\beta$ value would by ~14.4 and the origin time ts ~130 Myr.

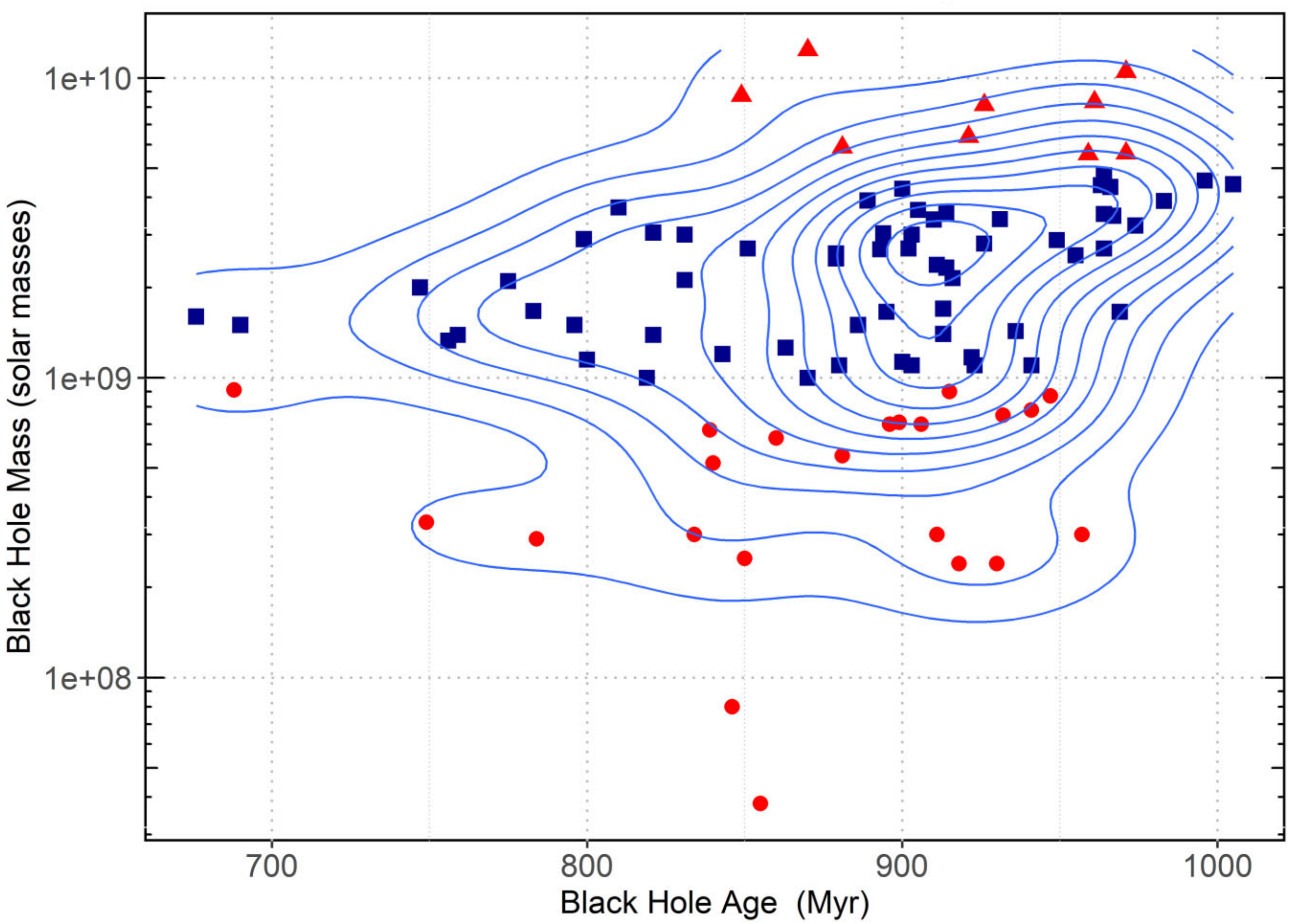

**Figure 1.** Mass $M_{\rm BH}$ versus age $t$ for 91 SMBHs are listed in the Appendix. Density contours are shown. The average reported 1$\sigma$ uncertainty in $M_{\rm BH}$ is ~0.11 dex. The 60 BHs denoted by squares share a common feature: their masses are within a factor of ~2 of 2.5 × 10$^9$ M$_\odot$, but their ages span ~325 Myr of cosmic time. Those in triangles nominally have $M_{\rm BH}$ > 5 × 10$^9$ M$_\odot$, and those in circles have $M_{\rm BH}$ < 10$^9$ M$_\odot$. The two BHs GNz11 ($t$ ~ 434 Myr; $M_{\rm BH}$ ~ 1.510$^6$ M$_\odot$) and CEERS_19 ($t$ ~ 568 Myr; $M_{\rm BH}$ ~9 × 10$^6$ M$_\odot$) belong to the group with circles and fall to the left and below the bottom left corner of the figure far from the rest of the BHs.

Substituting these values of $\beta$ and ts in equation (3), we get equation (4A) where age $t$ is in Myr. And using the approximation $1/t \propto (1+z)^{3/2}$ for high $z$ (Bergström & Goober 2006), we can rewrite equation (4A) expressing a BH's $M_{\rm BH}$ as a function of $z$ as in equation (4B). Note that equation (4A) and (4B) may yield slightly different results because of the approximation. Note also that equation (4) does not depend on any unjustified material assumption or ad hoc data selection.

$$M_{\rm BH} = {\rm Ms} \exp\left[14.6\,(t-100)\,/t\right], \tag{4A}$$

$$M_{\rm BH} = {\rm Ms} \exp 14.6\left[1-(1+z)^{3/2}/(1+30)^{3/2}\right]. \tag{4B}$$

## 5. APPLICATION TO HIGH-Z SMBHS: LIMITS ON SEED MASS

Fig. 2 shows the seed mass Ms versus age $t$ distribution for the 91 BHs in Fig. 1 predicted using equation (4). Figs 1 and 2 share identical $M_{\rm BH}$ symbols. Remarkably, the predicted Ms distribution in Fig. 2 closely mimics the observed $M_{\rm BH}$ distribution in Fig. 1. The 60 blue squares having similar $M_{\rm BH}$ have markedly similar Ms; the red circles having smaller $M_{\rm BH}$ have correspondingly smaller Ms; the red triangles with the largest $M_{\rm BH}$ have the largest Ms; and the BHs range in $M_{\rm BH}$ over ~2.5 orders of magnitude and so do their Ms. Fig. 3 shows the predicted seed mass Ms versus BH mass $M_{\rm BH}$ for all 93 BHs including GNz11 and CEERS_1019, the two red dots in the lower left corner of the figure. The symbols are the same as those in Figs 1 and 2. The largest BH in Fig. 3 (#31 in Table A1) requires a seed mass Ms = $(3 \pm 1) \times 10^4\,{\rm M}_{\odot}$ for a $2\sigma$ uncertainty in its mass. And those with masses 1–5 $\times\, 10^9\,{\rm M}_{\odot}$ (blue squares in Fig. 3) require seeds $> 10^3\,{\rm M}_{\odot}$ but $< 10^4\,{\rm M}_{\odot}$. Of the 93 BHs, GNz11, and CEERS_1019 are the smallest, requiring Ms $\sim$20 and $\sim$53 ${\rm M}_{\odot}$ respectively, or a few to several tens of solar masses. Additionally, equation (4) predicts that a seed of $\sim$418 ${\rm M}_{\odot}$ formed at ts = 100 Myr accounts for the mass of UHZ1 ($z = 10.1$) estimated to be $\sim 4 \times 10^7\,{\rm M}_{\odot}$ by Natarajan et al. (2024), who concluded that UHZ1 grew from a heavy seed ($\sim 10^4\,{\rm M}_{\odot}$) formed at the cosmic age of $\sim$300 Myr ($z \sim 14$) that was probably a DCBH.

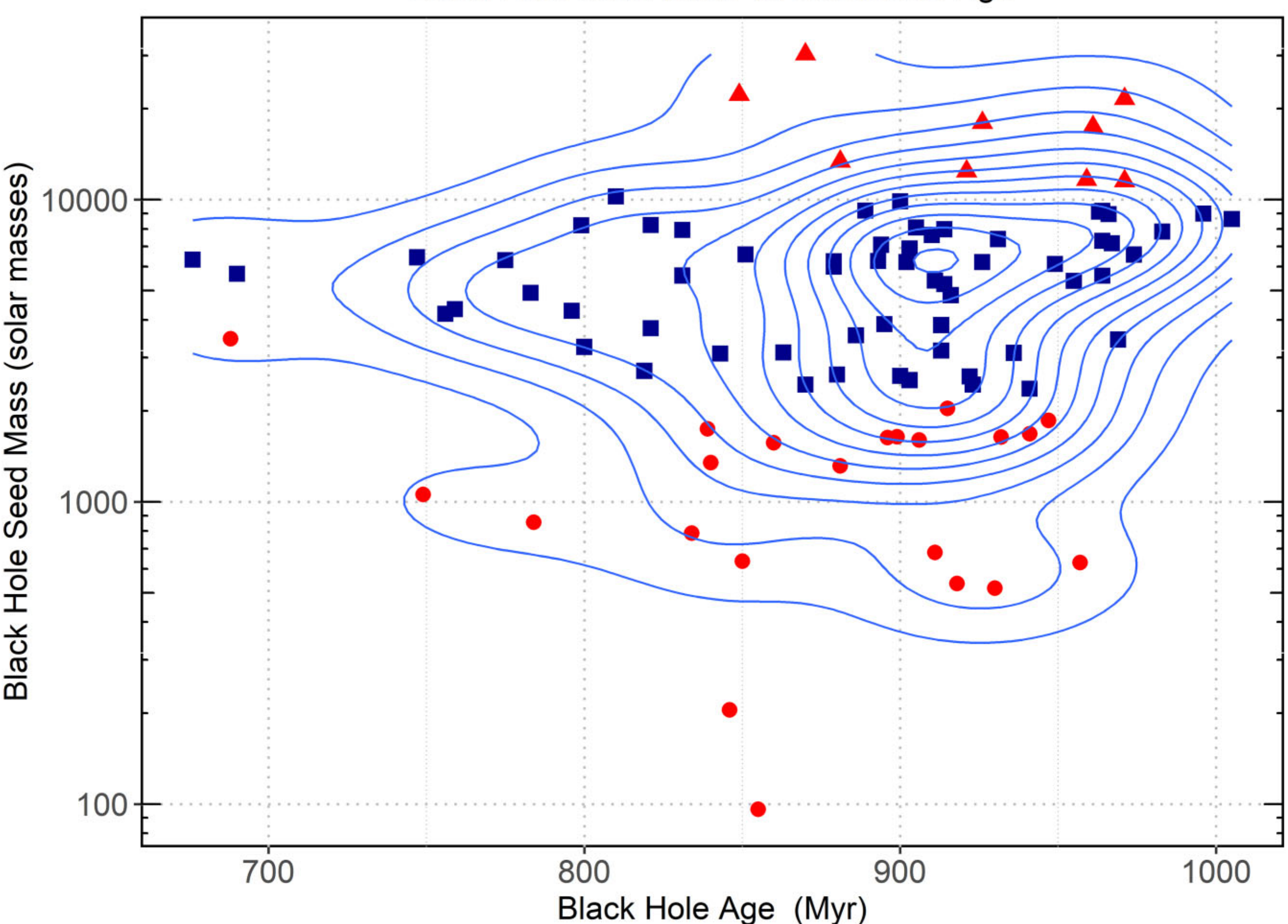

**Figure 2.** Seed mass Ms versus age *t* predicted by equation (4) for the BHs in Fig. 1. The symbols are the same as in Fig. 1, differentiating BHs based on their $M_{\rm BH}$. The distribution of Ms mimics the distribution of $M_{\rm BH}$ in Fig. 1, and so do the density contours.

Furthermore, we can place upper and lower limits on the mass of seeds that formed at $z \sim 30$. The largest high-$z$ BH (#31, Table A1) was discovered $\sim$10 yr ago, and the second largest (#59) more than two decades ago. In all likelihood, these BHs represent an upper limit on the size of SMBHs in the early universe. If so, we can conclude that the largest seeds formed at $z \sim 30$ did not exceed $(3 \pm 1) \times 10^4\,{\rm M}_{\odot}$, the seed mass required for the largest of the 93 SMBHs. Conversely, while we cannot place a strict lower limit on the mass of seeds formed at $z = 30$, we can conclude that it has to be $< 20\,{\rm M}_{\odot}$ based on the Ms required for GNz11, the smallest of the 93 high-$z$ SMBHs. Lastly, we note that equation (4A) predicts that the maximum size a BH can achieve via luminous accretion depends on its seed mass amounting to Ms times Exp14.6 or $\sim 2.2 \times 10^6$ Ms, which translates into $(6.6 \pm 2.2) \times 10^{10}\,{\rm M}_{\odot}$ for the empirically determined upper limit of Ms = $(3 \pm 1) \times 10^4\,{\rm M}_{\odot}$, in excellent agreement with a probable theoretical limit of $\sim 5 \times 10^{10}\,{\rm M}_{\odot}$ proposed by King (2012).

## 6. APPLICATION TO LOWER *Z* SMBHS: SEED CLASSIFICATION

We applied equation (4B) to the 132 446 AGN at $z < 2.4$ (Kozlowski, 2017) to test its validity and derive the mass distribution of the seeds. Furthermore, to assess the precision with which equation (4) predicts the seed mass, we applied it to a test group of 10 BHs with similar masses (1.51–1.77 $\times\, 10^{10}\,{\rm M}_{\odot}$) but different ages spread over $\sim$ a billion years from 2900–4204 Myr. The results show that these BHs had similar seed masses Ms within $\pm$ 0.3 per cent, in agreement with the inference drawn earlier from the distribution of high-$z$ BHs that the 60 BHs (blue squares, Fig. 1) of different ages within a narrow age range but similar masses must have had similar-mass seeds.

The resulting Ms for the 132 446 AGNs are sorted into narrow bins, and the number in each mass bin is shown in Table 1. In most cases, the uncertainty in $M_{\rm BH}$ is unknown but is of little importance except when there are relatively few seeds in a bin. Of the total population of BHs, 540 BHs have $\geq 10^{10}\,{\rm M}_{\odot}$, and none of their predicted Ms

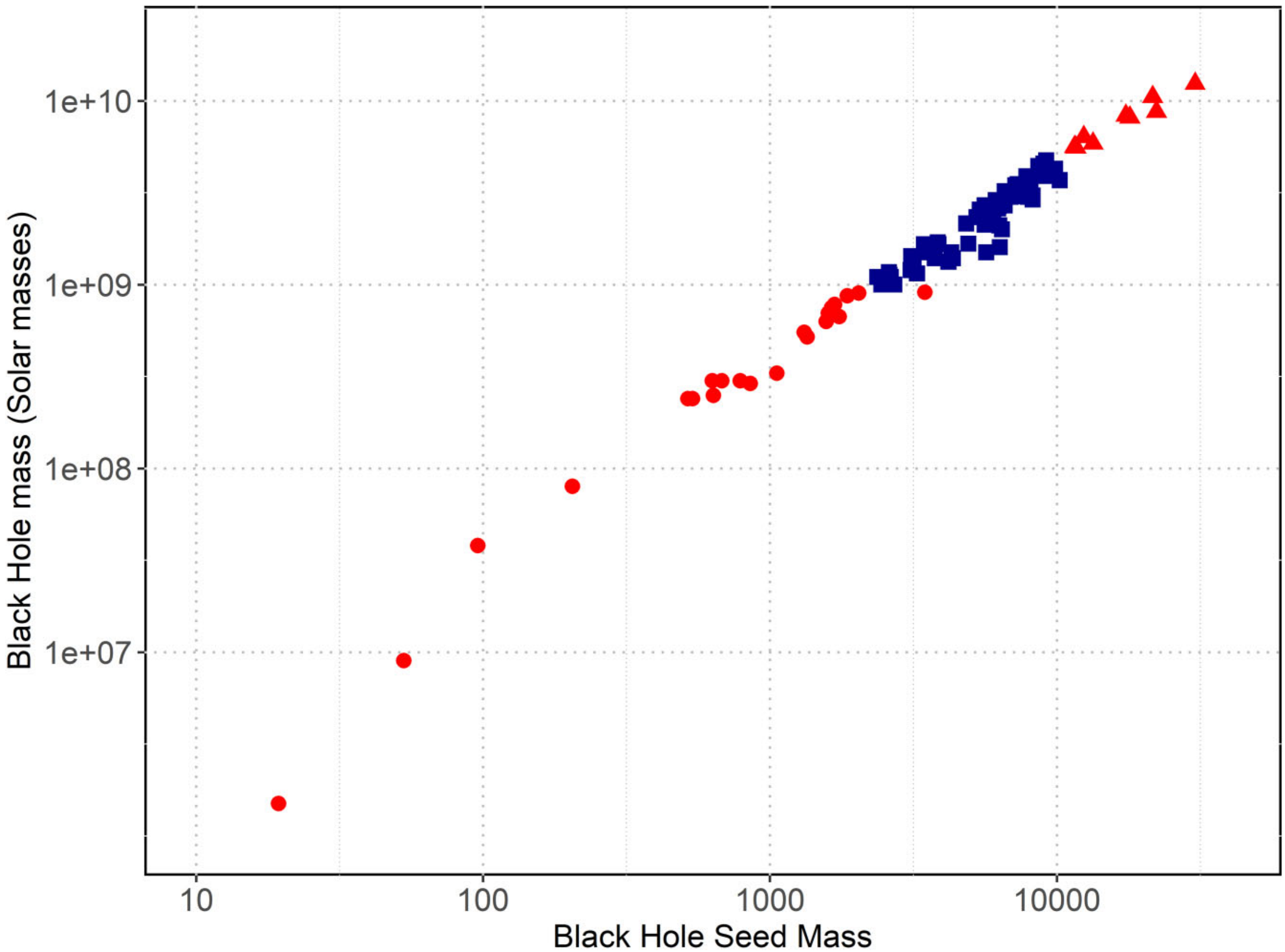

**Figure 3.** Seed mass Ms versus BH mass $M_{\rm BH}$ on a log–log scale for the 91 BHs in Fig. 1 and GNz11 and CEERS_1019 ($M_{\rm BH} < 1e+07 {\rm M}_{\odot}$, bottom left corner). The symbols are identical to those in Figs 1 and 2. The Ms for GNz11 and CEERS_1019 are ~20 and ~53 $\rm M_{\odot}$, respectively. The Ms for BHs with $M_{\rm BH}$ 1–5 × $10^9$ $\rm M_{\odot}$ (squares) have $10^3$ < Ms < $10^4$ $\rm M_{\odot}$, and the Ms for the largest BH is ~3 × $10^4$ $\rm M_{\odot}$.

**Table 1.** Seed counts.

| Ms | Count | Ms | Count | Ms | Count | Ms | Count |
|---|---|---|---|---|---|---|---|
| 5–10 | 88 | 200–250 | 8473 | 800–850 | 1 963 | 1400–1450 | 745 |
| 10–20 | 1014 | 250–300 | 7063 | 850–900 | 1 913 | 1450–1500 | 703 |
| 20–30 | 2394 | 300–350 | 6063 | 900–950 | 1 721 | 1500–1800 | 3383 |
| 30–40 | 3332 | 350–400 | 5241 | 950–1000 | 1 528 | 1800–2200 | 2775 |
| 40–50 | 3605 | 400–450 | 4700 | 1000–1050 | 1 419 | 2200–2700 | 2217 |
| 50–60 | 3595 | 450–500 | 4064 | 1050–1100 | 1 231 | 2700–3400 | 1747 |
| 60–70 | 3481 | 500–550 | 3728 | 1100–1150 | 1 161 | 3400–4200 | 1051 |
| 70–80 | 3261 | 550–600 | 3226 | 1150–1200 | 1 081 | 4200–6400 | 1097 |
| 80–90 | 3110 | 600–650 | 2910 | 1200–1250 | 944 | 6400–10 000 | 462 |
| 90–100 | 2979 | 650–700 | 2672 | 1250–1300 | 926 | 10 000–14 000 | 144 |
| 100–150 | 12 739 | 700–750 | 2385 | 1300–1350 | 843 | 14 000–30 000 | 61 |
| 150–200 | 10 159 | 750–800 | 2217 | 1350–1400 | 829 | >30 000 | 5 |

*Notes.* Ms seed mass (solar mass). The count is the number of seeds in a bin. Bin size varies.

exceed the preceding empirically established upper limit of (3 ± 1) × $10^4$ $\rm M_{\odot}$ except possibly in five cases (Table 1) and that too by a factor of < 1.7 in the worst case. However, the uncertainty in the $M_{\rm BH}$ of these five BHs is unknown. In particular, we note that the predicted Ms for TON 618 at $z = 2.219$ and 4.07 × $10^{10}$ $\rm M_{\odot}$ (Xue Ge et al. 2019), which is often cited as the most massive BH observed to date, is identical to that of the largest high-$z$ BH (#31). At the lower end of the Ms spectrum, the results (Table 1) show that 88 BHs have predicted *M*s between 5 and 10 $\rm M_{\odot}$ and ~1000 between 10 and 20 $\rm M_{\odot}$ in agreement with the lower limit of Ms expected from the high-$z$ data. This agreement on the upper and lower limits on the size of seeds deduced from 2 entirely different sets of data covering different cosmic periods cannot simply be fortuitous and testifies to the validity of equation (4).

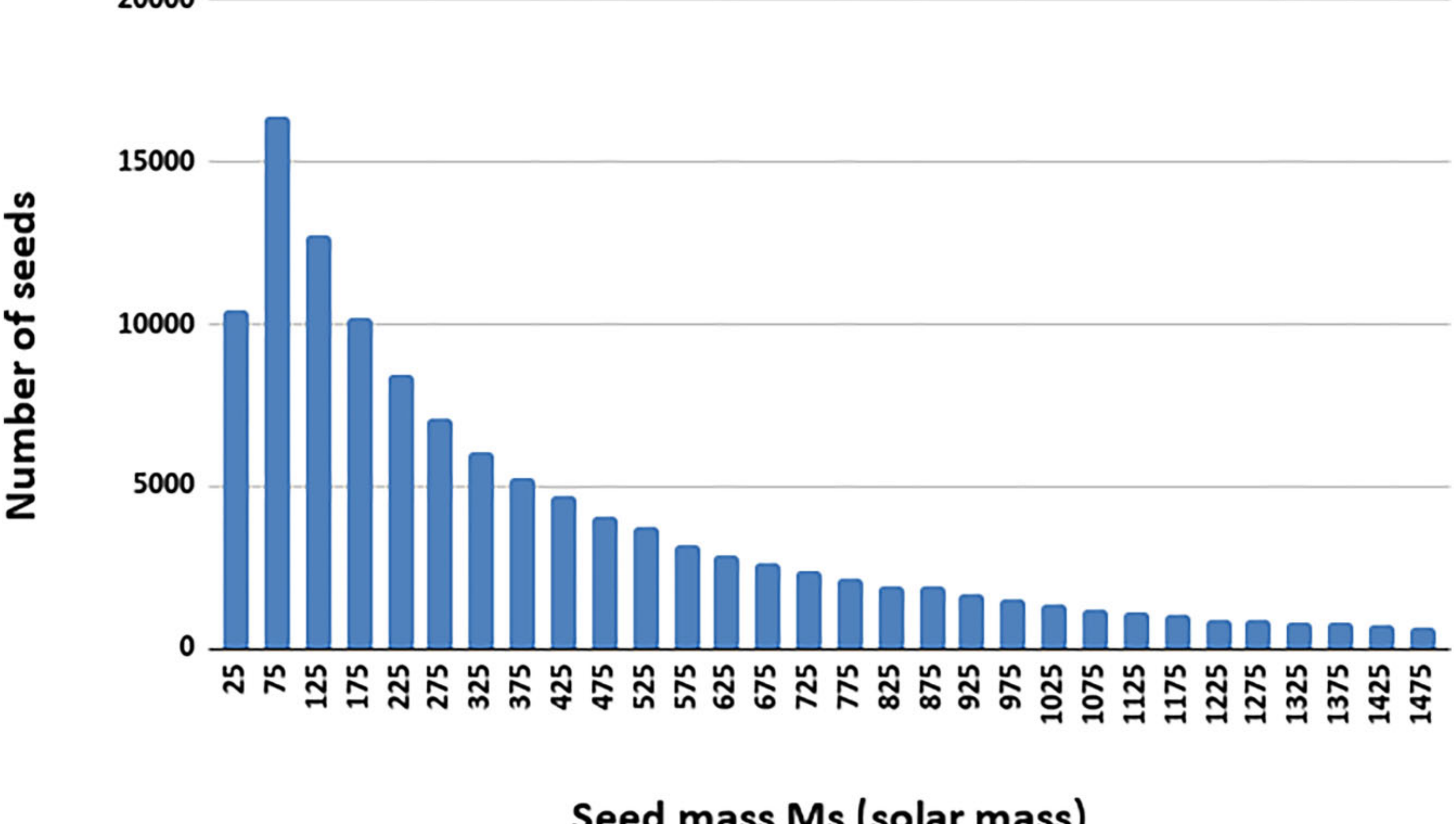

**Figure 4.** Mass distribution of 90 per cent of seeds of 132 446 SMBHs. Each bin has a width of 50 $M_\odot$ (solar masses) and is plotted at its central value. The distribution of the rest at Ms $> 1500\,M_\odot$, not plotted because of their small numbers, is given in Table 1. Note the initial increase in seed counts before the monotonic decrease, the details of which are shown in Fig. 5 and the significance of which is discussed in the text.

The seed counts in Table 1 indicate that ∼90 per cent of the seeds have Ms $\leq 1.5 \times 10^3\,M_\odot$. The histogram in Fig. 4 shows their mass distribution, where each bin has the same size of 50 $M_\odot$ and is identified by the central value of Ms in the bin. The remaining 10 per cent are not plotted because of their small numbers but follow the same pattern, decreasing asymptotically towards zero at Ms $\sim 3 \times 10^4\,M_\odot$, the upper limit of seed mass. Table 1, however, gives the number of seeds in different bin sizes for the remaining 10 per cent of AGN that amount to less than the number in the single bin of 50–100 $M_\odot$ centred at Ms = 75 $M_\odot$ in Fig. 4. The seeds in Fig. 4 range in Ms from a low of $\sim 5\,M_\odot$ to a high of $\sim 1.5 \times 10^3\,M_\odot$ and can be designated as light to intermediate-size seeds based on their universally accepted size classification They constitute an overwhelming majority of the seeds. In contrast, seeds $\geq 10^4\,M_\odot$ or heavy seeds constitute a small number totaling ∼210 or ∼0.16 per cent of the seed population (see Table 1). Furthermore, the distribution in Fig. 4 shows a monotonic systematic decrease in the number of seeds without any hiatus or break, preceded by an initial increase in seed counts.

Rather than reviewing the numerous simulations of BH seed formation reported in the literature, we shall attempt to identify those that best match the distribution of seeds in Fig. 4 and Table 1. Light seeds are thought to have formed from the collapse of massive metal-free first or Pop III stars (Madau & Rees 2001; Johnson & Bromm, 2007). Our finding that the seeds formed at $z \sim 30$ is consistent with the notion that the first stars formed at $z \sim 30$ (e.g. Couchman & Rees 1986). Their stellar mass functions derived from simulations range from $< 10\,M_\odot$ to a few cases of $\sim 1000\,M_\odot$ (e.g. Hirano et al. 2014; Hosokawa et al. 2016; Stacy et al. 2016). Latif & Ferrara (2016) reviewed these simulations and concluded that the results suggest that the typical mass of Pop III stars is $\sim 100\,M_\odot$.

Hydrodynamical simulations by Hirano et al. (2014, henceforth H14) indicate that the stellar mass function depends on the protostars's rate of accretion. Their simulations show that protostars accreting at the lowest rate result in general in smaller but more numerous stars than those accreting at higher rates. The upper panel in Fig. 5 shows the mass distribution of first stars copied with permission from H14, where histograms in red, blue and black represent stars formed from protostars accreting respectively at the lowest, intermediate, and the highest rate. For comparison, the lower panel in Fig. 5 shows the mass distribution of BH seeds from Table 1 (this study). First, we note that the overwhelming majority of first stars (upper panel) have masses $< 400\,M_\odot$; only eight stars are larger and thinly spaced over ∼400–1600 $M_\odot$. There are several similarities between the two histograms. In both, the number of objects initially increases, reaches a maximum or a plateau, and steadily decreases after that. Secondly, the number of stars initially increases exponentially by an order of magnitude from 1 to a maximum of 11, and so do the number of BH seeds from ∼88 to ∼3 600 (see Table 1). In both cases, the peak is reached at or near the bin with 40 solar masses. Clearly, the predicted mass distribution of light seeds (this study) thought to be the remnants of first stars; resembles the observed mass distribution of first stars in hydrodynamic simulations by H14; especially taking into consideration that the upper panel represents solely the distribution of first stars, whereas the lower panel probably represents the composite distribution of Pop III remnants and seed BHs formed by their mergers as discussed later. However, a notable difference between the two histograms is that there is a marked hiatus in the number of stars (upper panel) immediately following the maximum, but there is no such hiatus in the number of seed BHs (lower panel). Later, we provide an explanation reconciling this dichotomy.

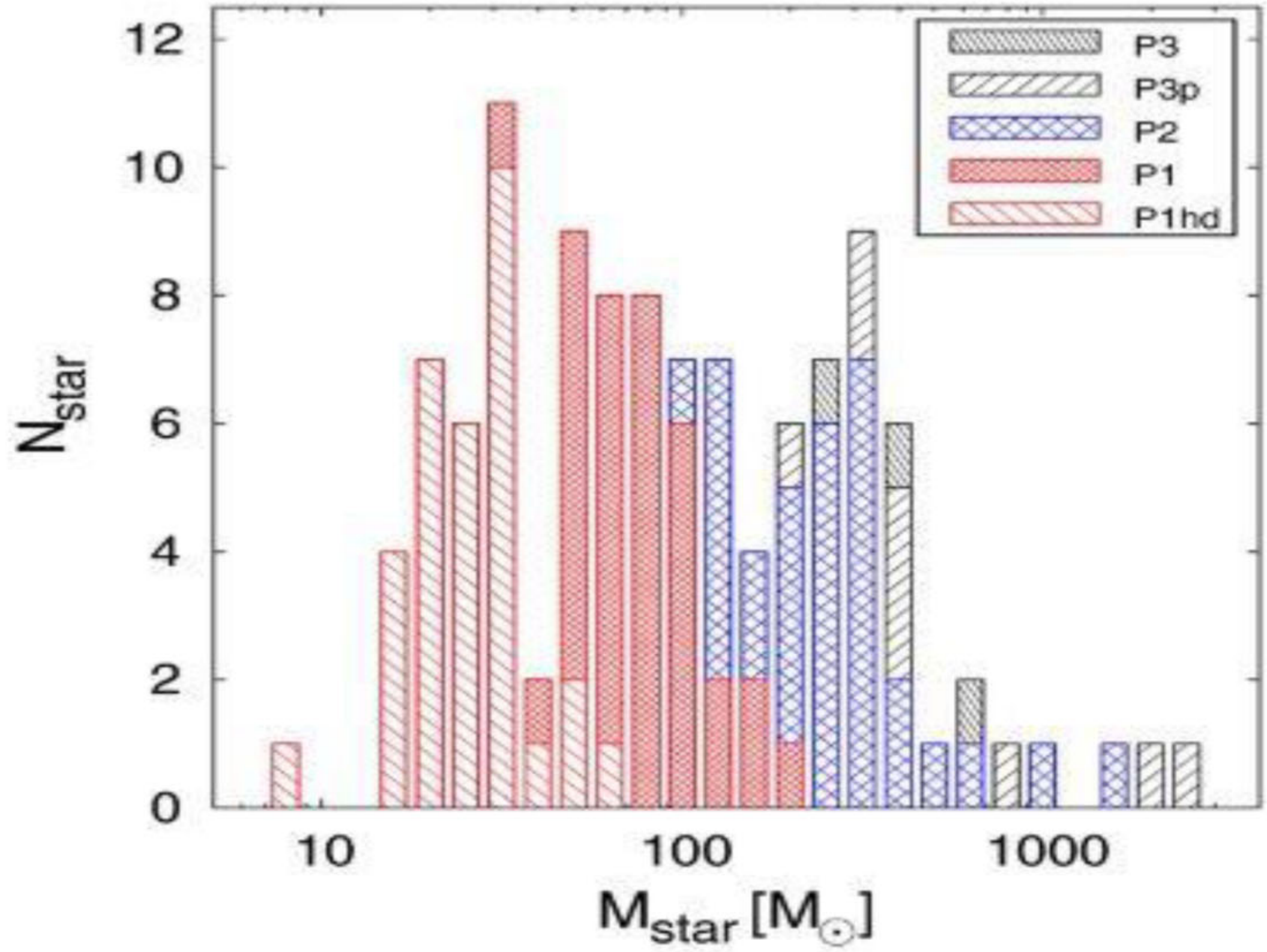

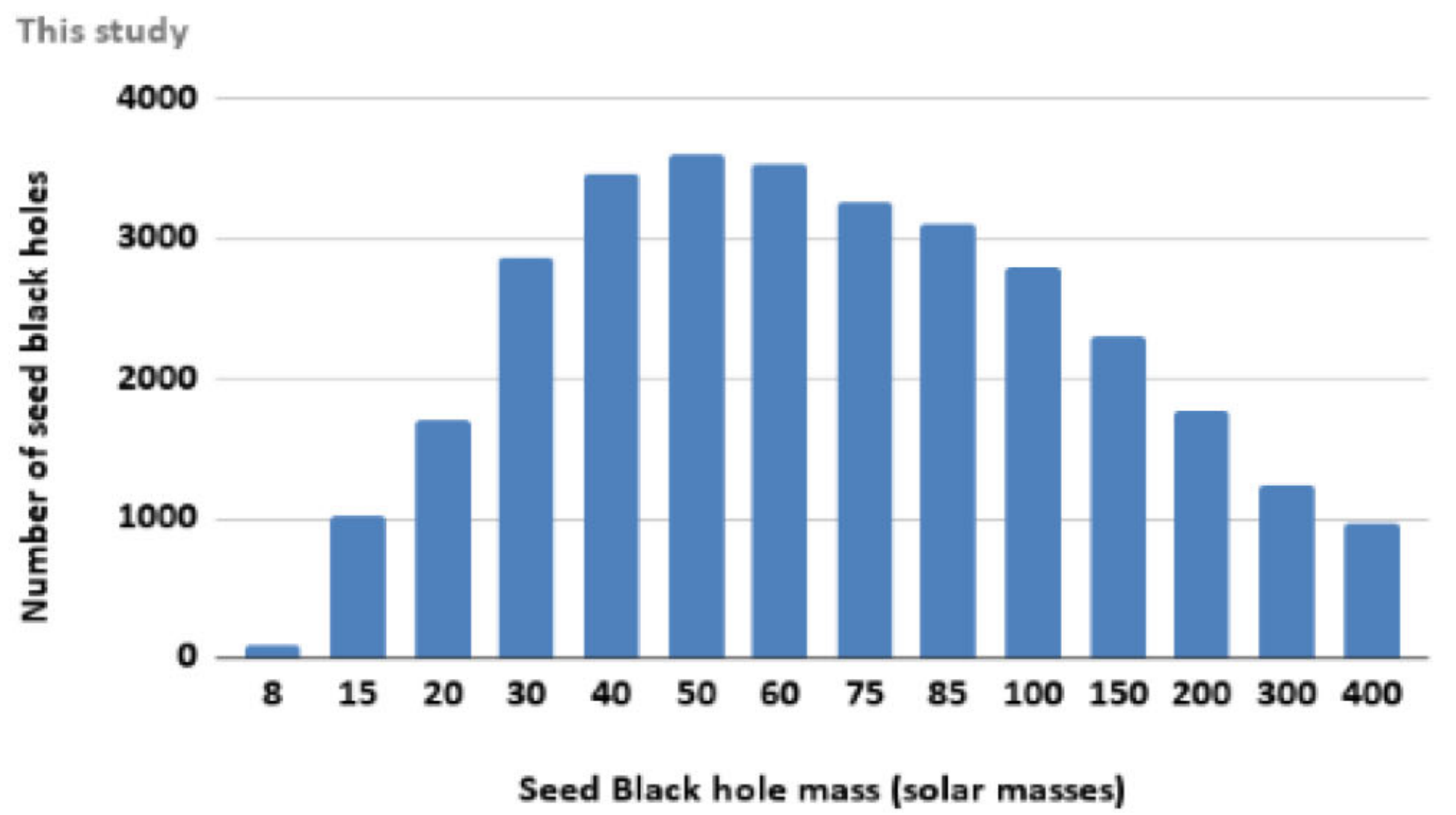

**Figure 5.** Mass distribution of first stars (upper panel from H14) and BHs seeds (this study, lower panel). The different colours and symbols in the upper panel represent protostars accreting at different rates. The bins in the lower histograms have a width of 10 solar masses and each bar is plotted at the central value of the bin.

In contrast to light seeds, intermediate-size seeds are thought to have formed either via runaway collisions of stars in dense stellar clusters (Portegies et al. 2004; Freitag et al. 2006; Mapelli, 2016) or by the hierarchical merger of BHs in stellar clusters (Davies et al. 2011; Lupi et al. 2014; Giersz et al. 2015). Most likely, there is an overlap in the sizes of light and intermediate seeds. There is, however, no decipherable change, hiatus, or break in the asymptotic decrease in the number of seeds as seed mass Ms increases in Fig. 4. Thus, it is difficult, if not impossible, to define a strict Ms boundary in Fig. 4 between the light and intermediate seeds. Devecchi et al. (2012) performed simulations simultaneously investigating the formation of Pop III remnants and BHs via runaway collisions in nuclear star clusters. Their results show that the BHs formed via runaway collisions decrease in number by a factor of ∼5 as BH mass increases from ∼400 to ∼3000 $M_\odot$ qualitatively in agreement with the results of this study (see Table 1). There is, however, a problem with the timing of formation of BH seeds by runaway collision of stars. While their results indicate that Pop III remnants formed around $z = 30$ in agreement with this study, the larger BHs formed much later starting at $z \sim 15$ in contradiction to the results of this study that exclude the possibility of BH seeds forming at $z \sim 15$ or later (see the discussion in the section on the derivation of empirical relations).

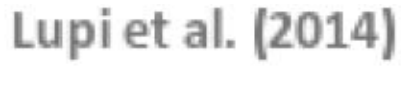

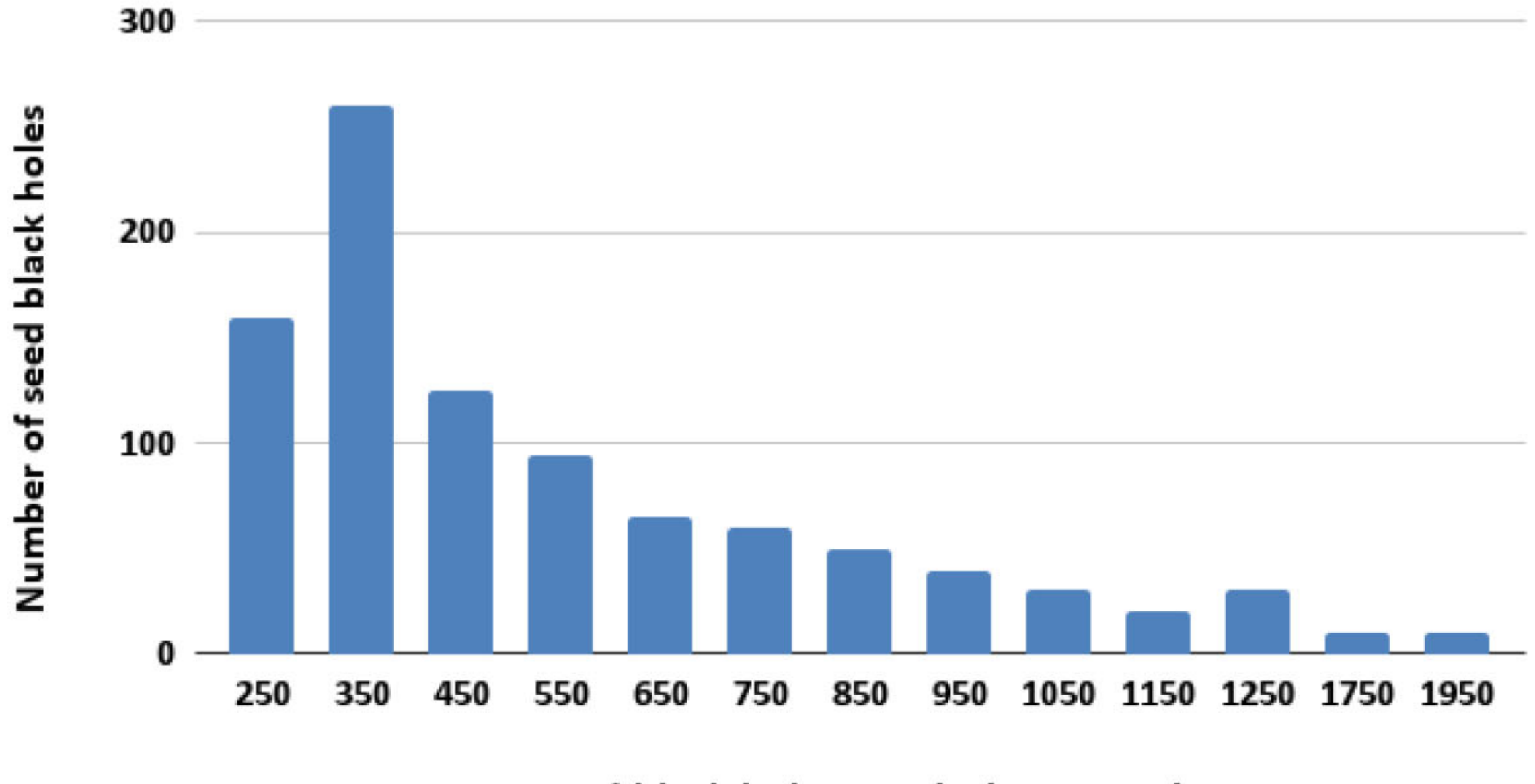

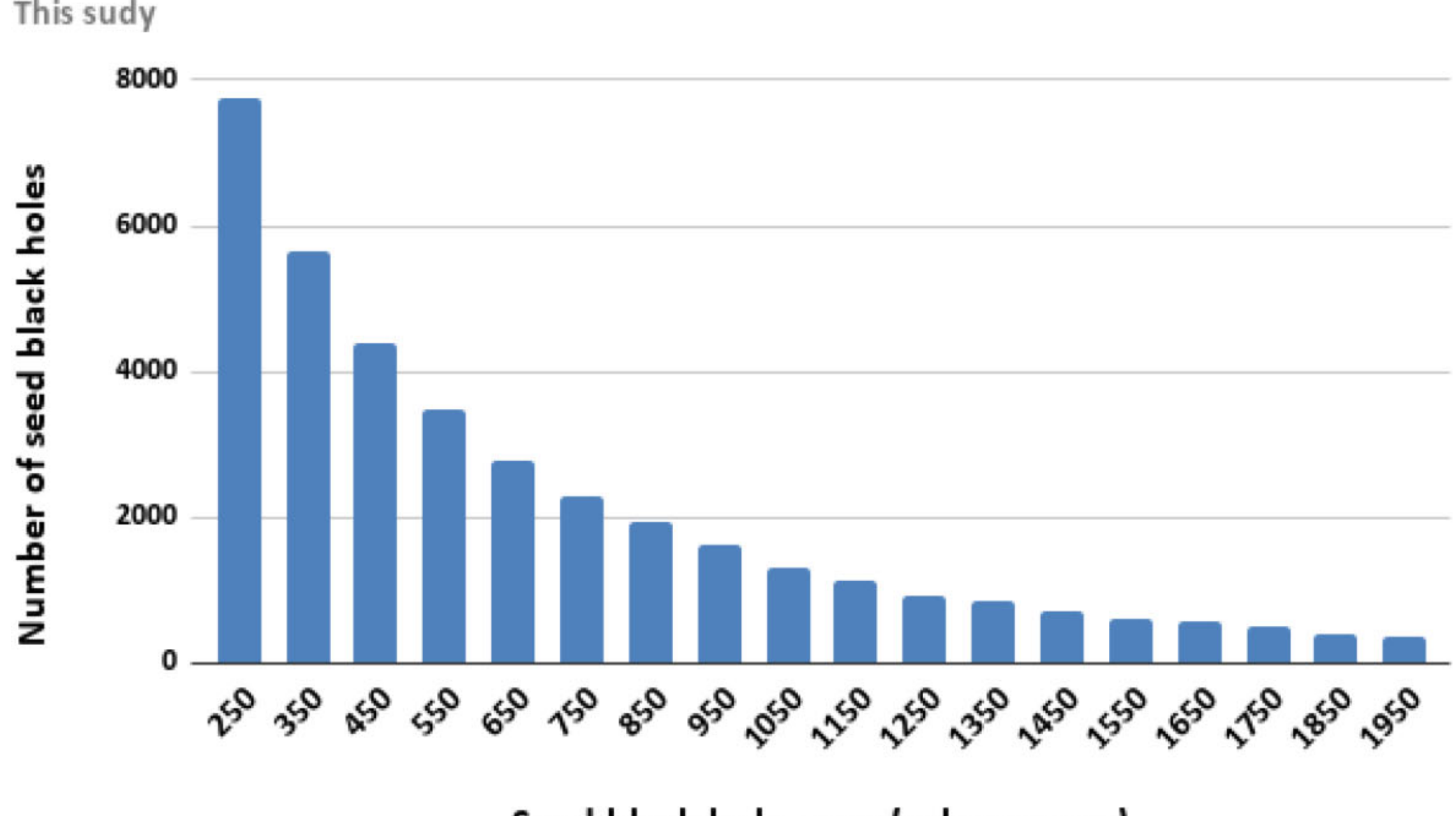

**Figure 6.** Comparison of mass distribution of BH seeds formed by GIRM or the gas-induced runaway merger of BHs (Lupi et al. 2014) and BH seeds in Table 1 (this study). Note that each bin in the lower panel has a width of 50 $M_{\odot}$ (solar masses) and is plotted on its central value.

On the other hand, Lupi et al. (2014) explored gas-induced runaway merger of BHs dubbed the GIRM model, following the Davies et al. (2011) prescription of hierarchical growth of BHs in dense stellar clusters. The number of BHs formed depends upon the value of a contraction parameter $\xi$ that provides a measure of contraction of the star cluster, and they performed simulations for three different values $\xi$. Fig. 6 (upper panel) shows the resulting mass distribution of BHs, adopted from Fig. 7 in Lupi et al. (2014). In comparison, the lower panel in Fig. 6 shows the distribution of BH seeds from this study. There are remarkable similarities in the two distributions. In both cases, the number of seeds systematically decreases as seed mass increases from ~350 to ~2000 $M_{\odot}$. Note also, that the magnitude of the decrease in the number of seeds from the peak to the last bin is comparable in both distributions. In the upper panel, it is by a factor of ~26 and by a factor of ~21 in the lower panel. The only apparent difference seems to be that in the upper panel the maximum number of BHs is at the bin with 350 $M_\odot$, whereas in the lower panel it is at 250 $M_\odot$. Most of the BHs in their simulations formed at $z < 20$, which could potentially be problematic in view of the results of this study. Lupi et al. (2014), however, point out that GIRM requires some degree of metal pollution of the intergalactic medium from the explosions of the Pop III stars and that the 'GIRM channel does not pose any constraint on the level of the metallicity of the parent halo, and hence on the time of formation, provided Pop III stars have enhanced the metallicity above a threshold'. If so, this mechanism and that for the formation of light seeds from the collapse of first stars could together account for at least 94 per cent of the seed population in Table 1. This mechanism has the advantage that the Pop III remnants could provide the degree of metal pollution required in the simulations. Alternatively, it is likely that the formation of intermediate-size BHs could have resulted from the merger of Pop III remnants; which would explain why there is no hiatus or break in the mass distribution of seeds in Fig. 4 and Table 1 and why intermediate-size follow the same asymptotically decreasing trend as the light seeds. Furthermore, the probability of forming larger seeds from a given population of stellar seeds would systematically decrease as seed mass increases, consistent with the systematic decrease in seed counts as Ms increases observed in Fig. 4 and Table 1.

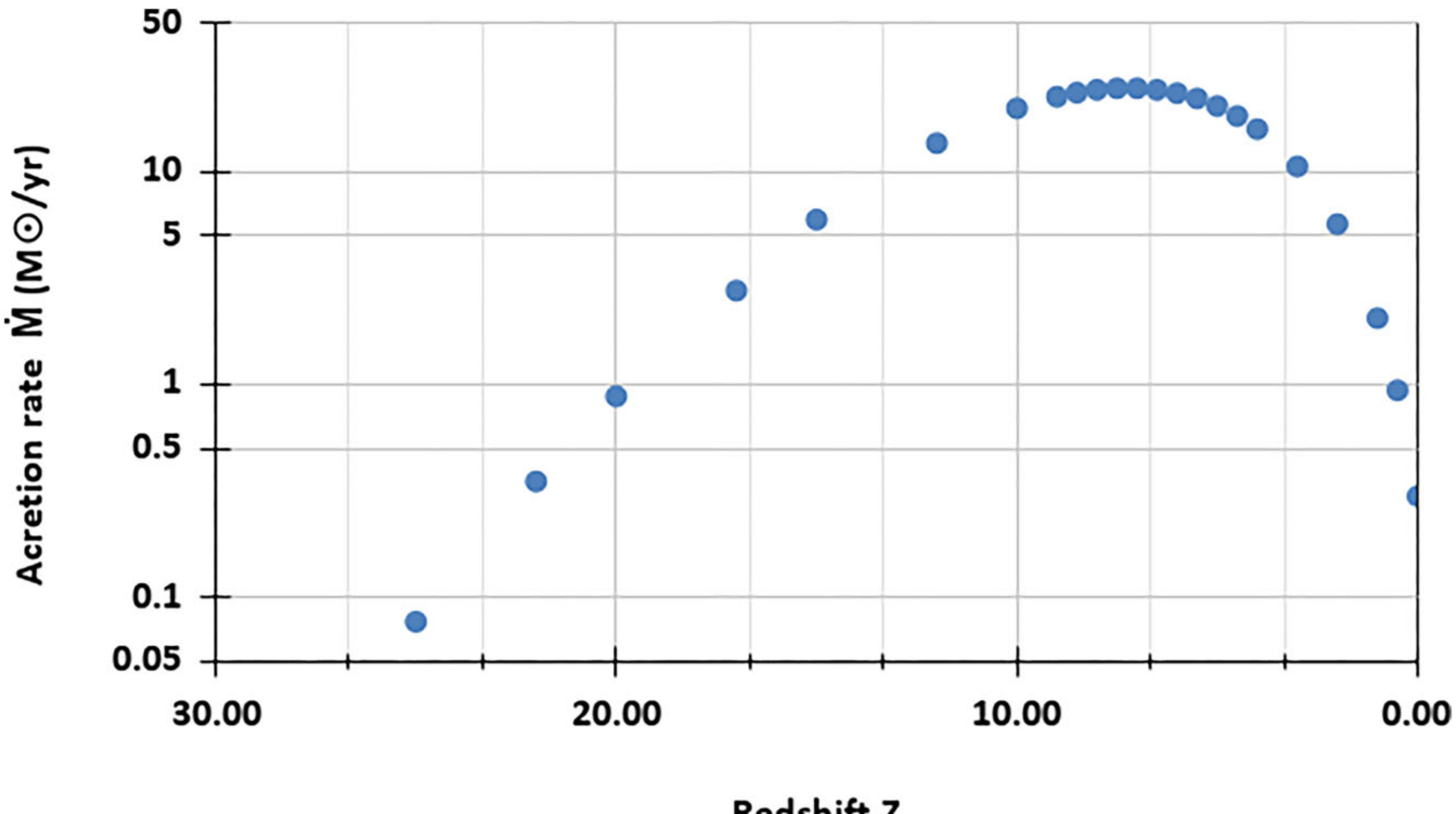

**Figure 7.** Shows the instantaneous accretion rate $\dot{M}$ for the largest high-$z$ SMBH (#31, Table A1) from its inception as a seed of $3 \times 10^4$ $M_\odot$ at $z \sim 30$ based on equation (6). Initially, $\dot{M}$ increases exponentially, reaches a broad peak between $z$ ~8.5 and 6 with a maximum at $z \sim 7$, and steadily decreases after that towards $z = 0$. Note that the timing of the broad peak from $z = 8.5$–6 coincides with the epoch during which reionization is thought to have occurred. Note also that $\dot{M}$ scales as seed mass $M$s.

Based on analyses of the mass distributions in Figs 5 and 6, we can now classify the seeds as per their masses. We noted that the vast majority of first stars in Fig. 5 have masses < 400 $M_\odot$ and BHs formed by hierarchical merger of seed BHs in Fig. 6 have masses > 350 $M_\odot$ Hence, we can reasonably classify BH seeds < 350 $M_\odot$ as light seeds predominantly formed from the collapse of first stars and those between 350 and 2000 $M_\odot$ as predominantly intermediate size assembled by the merger of stellar-mass seeds while recognizing that there is probably an overlap between the classes. Based on this working classification, we conclude that of the 132 446 seeds in Table 1, ~54 per cent are light, and ~40 per cent are intermediate size. The remaining ~6 per cent have Ms between ~$2 \times 10^3$ and $3 \times 10^4$ $M_\odot$ (dubbed heavier seeds in contrast to heavy), of which only 210 (~0.16 per cent) have Ms $\geq 10^4 M_\odot$ that could strictly be classified as heavy seeds. Moreover, these relatively few heavy seeds have Ms $\leq 3 \times 10^4 M_\odot$ or at the lower end of the expected mass spectrum of $10^4$–$10^5$ $M_\odot$ (see Volonteri et al. 2021) for heavy seeds thought to have formed by the DCBH mechanism. The implication is that the DCBH mechanism did not play an important role in seed formation. Moreover, the 'number of 'heavier' seeds decreases as Ms increases (see Table 1), following the same pattern as the light to intermediate-size seeds in Fig. 4, which suggests that the same mechanism responsible for forming intermediate-size seeds may also account for the rare formation of heavy seeds. Apropos, Davies et al. (2011) have proposed that seeds as large as $10^5$ $M_\odot$ can be formed by hierarchical growth of BHs in dense stellar clusters. We conclude, therefore, that the DCBH mechanism is not required while not closing the possibility that a relatively small number may have formed by the DCBH mechanism.

## 7. ACCRETION RATE, EDDINGTON RATIO, AND RADIATIVE EFFICIENCY

Having established the applicability of equation (4), we can use its derivative to gain insights into the growth history of SMBHs. Differentiating $M_{\rm BH}$ ($dM_{\rm BH}/dt$) in equation (4A), we get equation (5) expressing a BH's instantaneous accretion rate $\dot{M}$ ($M_\odot$ yr$^{-1}$) as a function of $M_{\rm BH}$ and $t$ (Myr) or as a function of $z$ using the approximation $1/t \propto (1+z)^{3/2}$ (Bergström & Goober 2006)

$$\dot{M}(\mathrm{M_\odot/yr}) = 14.6 \times 10^{-4} M_{\rm BH}/\mathrm{t}^2 \ \simeq 4.96 \times 10^{-12} M_{\rm BH}(1+z)^3. \quad (5)$$

Note that in equation (5), $\dot{M}/M_{\rm BH}$ or the accretion rate per unit BH mass is directly proportional to $(1 + z)^3$ and hence decreases as $z$ decreases, consistent with the requirement in the section on the theoretical basis that $\dot{M}/M_{\rm BH}$ decrease as $z$ decreases. Furthermore, in the Standard Cosmological Model, the ambient gas density scales as $(1+z)^3$, and hence the accretion rate $\dot{M}$ scales as the product of a BH's mass $M_{\rm BH}$ and the ambient gas density. Equation (5) can be applied to any actively accreting BH to ascertain its instantaneous accretion rate $\dot{M}$. For example, equation (5) yields $\dot{M} \sim 0.012\,\mathrm{M_\odot\,yr^{-1}}$ for GNz11 at $z = 10.6$, $\sim 24\,\mathrm{M_\odot\,yr^{-1}}$ for the largest (#31) high-$z$ SMBH at $z = 6.3$, and $\sim 6.6\,\mathrm{M_\odot\,yr^{-1}}$ for TON 618 at $z = 2.219$.

Unfortunately, there are no reported quantitative estimates of BH accretion rates supported by observational data that could be used to test equation (5). And estimates of a BH's accretion rate from its bolometric luminosity requires knowledge of its radiative efficiency; but the radiative efficiency has recently been shown (Aggarwal 2024) to be a function of both redshift and BH's mass, and using the standard value of 0.1 leads to highly erroneous results. However, the Bondi accretion rate $\dot{M}_{\rm B}$ is available for a few BHs in nearby galaxies (M87, NGC 3115, and NGC 1600) based on *Chandra* X-ray observations. One way to test equation (5) is that $\dot{\rm M}$ cannot be greater than $\dot{M}_{\rm B}$ because the latter is the rate of inflow of gas from a BH's Bondi radius, some of which may not reach the BH (see e.g. Yuan & Narayan, 2014) and some would inevitably be radiated away by the BH. Russell et al. (2015) obtained $\dot{M}_{\rm B} \sim 0.2\,\mathrm{M_\odot\,yr^{-1}}$ for M87 for $M_{\rm BH} = 6.6 \times 10^9\,\mathrm{M_\odot}$; the corresponding $\dot{M}$ value from equation (5) is $\sim 0.033\,\mathrm{M_\odot\,yr^{-1}}$. For NGC 3115, Wong et al. (2011) obtained $\dot{M}_{\rm B} \sim 0.022\,\mathrm{M_\odot\,yr^{-1}}$ for $M_{\rm BH} = 2 \times 10^9\,\mathrm{M_\odot}$; the corresponding $\dot{M} \sim 0.01\,\mathrm{M_\odot\,yr^{-1}}$. For NGC1600, Runge & Walker (2021) got $\dot{M}_{\rm B} \sim 0.15\,\mathrm{M_\odot\,yr^{-1}}$ for $M_{\rm BH} = 1.7 \times 10^{10}\,\mathrm{M_\odot}$; the corresponding $\dot{M} \sim 0.08\mathrm{M_\odot\,yr^{-1}}$. In addition, using observations of mass-losing stars, Quataert et al. (1999) got $\dot{M}_{\rm B} \geq 3 \times 10^{-5}\,\mathrm{M_\odot\,yr^{-1}}$ for Sagittarius A$*$ using $M_{\rm BH} = 2.6 \times 10^6\,\mathrm{M_\odot}$; which translates into $\dot{M}_{\rm B} \geq 5 \times 10^{-5}\,\mathrm{M_\odot\ yr^{-1}}$ for its current best estimate of $M_{\rm BH} = 4.15 \times 10^6\,\mathrm{M_\odot}$ (Gravity Collaboration 2019); the corresponding accretion rate from equation (5) is $\dot{M} \sim 2 \times 10^{-5}\,\mathrm{M_\odot\ yr^{-1}}$. In each case, the accretion rate $\dot{M}$ is smaller but comparable to the Bondi rate $\dot{M}_{\rm B.}$ While this comparison supports equation (5), we stress, that no conclusions should be drawn from this comparison as to how much of $\dot{M}_{\rm B}$ reaches a BH. It would require a thorough evaluation of any systematic errors in determining $\dot{M}_{\rm B}$ (see e.g. Wong et al. 2014); a task beyond the scope of this paper.

Moreover, by substituting $M_{\rm BH}$ in equation (5) with its expression in equation (4B), we get equation (6) expressing $\dot{M}$ as a function of a BH's seed mass Ms and redshift $z$.

$$\dot{M} = 4.96 \times 10^{-12}\mathrm{Ms}(1+z)^3 \exp 14.6\left[1-(1+z)^{3/2}/(1+30)^{3/2}\right]. \quad (6)$$

Using equation (6). one can infer the history of a BH's accretion rate from its inception at $z = 30$ to any redshift $z$ by plugging in equation (6) the BH's seed mass Ms inferred from equation (4B). Fig. 7 shows the accretion rate history of the largest high-$z$ SMBH (#31, Table A1), starting from a seed mass of $3 \times 10^4\,\mathrm{M_\odot}$ at $z = 30$. And since $\dot{M}$ scales as seed mass $M$s, the relative change in $\dot{M}$ with $z$ for any given Ms is the same as that illustrated in Fig. 7. Initially, $\dot{M}$ increases exponentially and reaches a broad peak between $z = 8.5$ and 6, beyond which it decreases slowly but monotonically towards $z = 0$. Note that the broad peak from $z = 8.5$–6, during which a BH has the highest accretion rate and hence likely the highest luminosity, incidentally coincides with the epoch during which reionization is thought to have occurred (e.g. Grazian et al. 2018). However, exploring to what extent BHs during this period (the mass function of which can be inferred using the results in Table 1 and equation 4) may have contributed to the reionization is beyond the current scope of this paper. Notwithstanding this sidebar, we note that two competing factors, namely the increase in a BH's mass or gravitational reach as it ages and the decrease in the ambient gas density as $z$ decreases, determine a BH's $\dot{M}$ at any instant of its life. Hence, the results in Fig. 7 suggest that initially the BH's mass or its gravitational reach increases faster than the decrease in the ambient gas density. The two competing factors reach parity near $z \sim 7$, after which the decline in gas density dominates over the gradual increase in the BH's mass and $\dot{\rm M}$ steadily decreases. For example, equation (4B) predicts that the seed of TON 618, one of the largest SMBHs, grew from $M$s $= 3.03 \times 10^4\,\mathrm{M_\odot}$ by $\sim 5.5$ orders of magnitude to $\sim 9.8 \times 10^9\,\mathrm{M_\odot}$ by $z = 7$ and after that by a factor of $\sim 4$ until its present $z = 2.219$. And equation (6) predicts that its accretion rate changed from $0.08\,\mathrm{M_\odot\,yr^{-1}}$ at $z = 25$ to $24.6\,\mathrm{M_\odot\,yr^{-1}}$ at $z = 7$ and $6.6\,\mathrm{M_\odot\,yr^{-1}}$ at $z = 2.219$.

Using equation (5), we can also define the Eddington ratio $\lambda$, or the ratio of a BH's bolometric luminosity $L_{\rm bol}$ to its Eddington luminosity ($L_{\rm EDD}$), as a function of $z$ and radiative efficiency $\varepsilon$. $L_{\rm bol} = (\dot{\rm M}c^2)\,\varepsilon/(1 - \varepsilon)$ where $c$ is the velocity of light and $\dot{M}$ the accretion rate; and $L_{\rm EDD} = 1.3 \times 10^{38}\,M_{\rm BH}$ in $\mathrm{erg\,s^{-1}}$ with $M_{\rm BH}$ in solar mass. Radiative efficiency $\varepsilon$ is conventionally defined relative to the mass inflow rate, such as the Bondi rate, and a BH's $\dot{M}$ is smaller by $(1 - \varepsilon)$. Thus, by substituting $\dot{M}$ from equation (5) into the definition of $L_{\rm bol}$, we get equation (7) expressing the Eddington ratio $\lambda$ as a function of a BH's redshift and radiative efficiency.

$$\lambda = 2.18 \times 10^{-3}(1+z)^3\varepsilon/(1-\varepsilon). \quad (7)$$

Equation (7) implies that $\lambda$ decreases as $z$ decreases, irrespective of whether $\varepsilon$ depends on $z$. This implication is validated by empirical evidence. Using Kozlowski's data for the tens of thousands of AGNs at $z < 2.4$. Aggarwal (2024) unambiguously showed that $\lambda$ decreases as $z$ decreases. Moreover, equation (7) implies that of 2 BHs at similar redshifts, the BH with a higher $\lambda$ is less efficient (higher $\varepsilon$) in accreting gases than the one with a lower $\lambda$. This implication is also consistent with the finding by Aggarwal (2024) that larger BHs have lower $\lambda$ and are more efficient than smaller BHs. And, using the Bondi prescription for spherically symmetric accretion and observational data for temperature and density profiles near BHs in galaxies M87, NGC 3115, and NGC 1600, Aggarwal (2024) derived a scaling relation for $\lambda$ identical in form to the above equation (7). Thus, the substantiations of the implications of equation (7) and the similarity between scaling relations for $\lambda$ derived from two different prescriptions and entirely different data sets are further evidence of the validity and applicability of equation (4) from which equation (7) is derived.

A BH's radiative efficiency $\varepsilon$ can be inferred from equation (7), knowing its $\lambda$ or bolometric luminosity. Unfortunately, determining $\lambda$ is prone to multiple errors arising from uncertainties in the BH's mass, its luminosity, and the correction factor used to get its bolometric luminosity. Hence, equation (7) should be used with caution. None the less, equation (7) implies that a BH's radiative efficiency increases as it ages or $z$ decreases as illustrated by the following example, in agreement with a recent finding by Aggarwal (2024). Shen et al. (2019) list 50 BHs, most of which have 1–$4 \times 10^9\,\mathrm{M_\odot}$ with redshifts close to 6 and a mean $\lambda \sim 0.32$, whereas a large group of similar-size BHs at $z \sim 1$ in Kozlowski's catalogue have a mean $\lambda = 0.03$; a decrease in $\lambda$ by a factor of $\sim$10 from $z = 6$ to 1. However, the corresponding decrease in the term $(1 + z)^3$ in equation (7) is $\sim$40 times. Hence, in equation (7), the term $\varepsilon/(1 - \varepsilon)$ must increase from $z = 6$ to 1 and so must $\varepsilon$ by a factor that depends upon the value of $\varepsilon$ at $z = 6$. Conversely, the implication is that $\varepsilon$ decreases as $z$ increases.

The vast majority of the values of $\lambda$ for the BHs in Table A1 and reported by Shen et al. (2019) and Kozlowski (2017) are $< 1$, all of which are at $z < 7.7$. GNz11 is a notable exception that deserves special attention because it is the highest-$z$ AGN observed so far and is inferred to be accreting by as much as 5.5 times the Eddington rate (Maiolino et al. 2024). If so, equation (7) predicts a radiative efficiency $\varepsilon$ of $\sim$0.59 that ostensibly is exceptionally high, especially since $\varepsilon$ has been shown to decrease as $z$ increases. And, it would imply that GNz11 is a relatively poor accreter, accreting only $\sim$41 per cent of the gas inflow. Interestingly, a lower $\lambda$ would make GNz11 a more efficient accreter. For example, $\lambda = 1$ would yield $\varepsilon = 0.227$, making GNz11 $\sim$2.6 times more efficient than if its $\lambda = 5.5$. Its $\lambda$ is probably highly overestimated. We note that Schneider et al. (2023) estimated a significantly lower $\lambda$ of 2–3 for it, and Bhatt et al. (2024) found that the probability of observing a BH at $z \sim 10-11$ accreting with $\lambda \sim 5.5$ in the volume surveyed by *JWST* is $< 0.2$ per cent.

Given that $\lambda$ is prone to large uncertainties and hence a poor predictor of whether a BH is accreting above the Eddington limit, we propose instead the following. As noted earlier, the growth efficiency parameter ɣ defined by equation (2) has a value of 1 for a BH accreting at the Eddington limit ($\lambda = 1$) with a radiative efficiency $\varepsilon = 0.1$ (its canonical value) and a duty cycle $\delta = 1$. For $\delta = 1$, a value of ɣ $>1$ implies that the BH is either accreting above the Eddington limit ($\lambda > 0.1$) or that its $\varepsilon < 0.1$. Substituting $\lambda$ from equation (7) into equation (2) (the definition of ɣ), one gets ɣ $\sim 2.4 \times 10^{-4}\delta\,(1 + z)^3$ for the instantaneous value of ɣ as opposed to its value averaged over a BH's lifespan in equation (1). Shankar et al. (2010) found that in their sample of AGNs, the duty cycle $\delta$ increased as $z$ increased reaching $\sim$0.9 at $z = 6$. It is likely, therefore, that $\delta \sim 1$ at even higher redshifts. Hence, at very high redshifts, the instantaneous value of ɣ is solely a function of the gas density. For $z > 15$, ɣ $>1$, and the implication is that either $\lambda > 1$ or $\varepsilon < 0.1$ at $z > 15$. Therefore, it is likely that during the first $\sim$150 Myr ($z \geq 15$) of its life, a BH either experienced super-Eddington accretion or its radiative efficiency was $< 0.1$. This finding is consistent with the suggestion by Wythe and Loeb (2012) and Pacucci et al. (2015) that super-Eddington accretion is possible when a BH is embedded in sufficiently dense gas that renders the radiation pressure less effective. For example, in the Standard Model, the ambient gas density at $z = 20$ is 27 times greater than at $z = 6$.

## 8. CONCLUSIONS

Prompted by insights derived from a deconstruction of the so-called Salpeter relation (equation 1–3), we analysed the mass versus age distribution of 91 high-$z$ SMBHs (Fig. 1) that resulted in the formulation of equation (4), the foundation on which the findings and conclusions of this paper are based. It describes a BH's mass $M_{\rm BH}$ as a function of its age $t$ or redshift $z$, from which the BH's seed mass Ms can be determined. It was extensively tested throughout the paper by verifying its implications and predictions. It, together with its derivates (equations 5–7), comprise a set of powerful tools to decipher the origins, growth, and properties of SMBHs.

We applied equation (4) to 93 high-$z$ ($>5.6$) and 132 446 AGNs at $z < 2.4$ listed by Kozlowski (2017). The results (Figs 2–4, and Table 1) show that the masses of the smallest to the largest actively accreting SMBHs observed to date are accounted for by seeds formed at $z \sim 30$ ranging in Ms from a low of $\sim 5\,{\rm M}_\odot$ to a maximum of $(3 \pm 1) \times 10^4\,{\rm M}_\odot$. In particular, the $M_{\rm BH}$ of GNz11, CEERS_1019, and UHZ1, the three highest redshift ($z = 8.7$–10.6) AGNs discovered recently, are accounted for by stellar-mass seeds ranging from a few tens to several hundred solar masses. Specifically, the results exclude the possibility that the seed of UHZ1 was heavy, presumably a DCBH, as postulated by Natarajan et al. (2024). Equation (4A) places an upper limit of $\sim$2, $2 \times 10^6$ Ms on the mass a seed can accrete via luminous accretion, which translates into $(6.6 \pm 2.2) \times 10^{10}\,{\rm M}_\odot$ for Ms $= (3 \pm 1) \times 10^4\,{\rm M}_\odot$ in agreement with a theoretical limit proposed by King (2012) and with the size of the largest SMBHs observed to date.

The mass distribution of seeds shown in Fig. 4 and Table 1 was analysed and compared with the simulated mass functions of first stars and intermediate-size BHs reported in the literature. Based on this comparative analysis, we classified the seed population into three broad categories depending on seed mass and the likely mechanism for its formation. We fully recognize that there is an overlap between the categories. Seeds $< 350\,{\rm M}_\odot$ were classified as light seeds predominantly formed from the collapse of massive metal-free first stars (Madau & Rees 2001; Johnson & Bromm 2007). Seeds ranging from 350–$2 \times 10^3\,{\rm M}_\odot$ were classified as intermediate size formed by the hierarchical growth via runaway mergers of BHs (Davies et al. 2011; Lupi et al. 2014). Light seeds constitute $\sim$54 per cent and intermediate size $\sim$ 40 per cent of the population in Table 1. The remaining $\sim$6 per cent raging in mass from $\sim 2 \times 10^3$ to $\sim 3 \times 10^4\,{\rm M}_\odot$ are dubbed heavier seeds in contrast to the classical heavy seeds ($10^4$–$10^5\,{\rm M}_\odot$) thought to be DCBHs. Of the heavier seeds, only a minuscule number (210) have $> 10^4\,{\rm M}_\odot$ and that too at the lower end of the presumed mass of DCBHs, which led us to conclude that the DCBH mechanism did not play a significant role and propose that the heavier seeds could also have formed via the merger of light to intermediate-size seeds.

Equation (5) gives a BH's instantaneous accretion rate as a function of its mass $M_{\rm BH}$ and age $t$ or redshift $z$. Equation (6) gives the same in terms of its seed mass Ms (inferred from equation 4) and $z$. Initially, $\dot{M}$ increases exponentially and reaches a broad plateau between $z = 8.5$–6, after which it decreases monotonically. Two factors, namely the increase in a BH's gravitational reach as its mass increases and the decrease in gas density as $z$ decreases, modulate the accretion rate as the BH ages. A BH's mass increases by $\sim$6 orders of magnitude in the first billion years and only by a factor of $\sim$4 in the next $\sim$12.8 billion years. Also, we noted that the timing ($z = 8.5$–6) of the maximum accretion rate in a BH's history coincides with the epoch when reionization is thought to have occurred.

Equation (7) expresses the Eddington ratio $\lambda$ as a function of a BH's $z$ and radiative efficiency $\varepsilon$. It implies that $\lambda$ decreases as $z$ decreases, but $\varepsilon$ increases as $z$ decreases; implications substantiated by unambiguous empirical evidence (see Aggarwal, (2024). Furthermore, a BH's radiative efficiency $\varepsilon$ can be determined using equation (7) from its $\lambda$. We, however, stressed that $\lambda$ is prone to large errors resulting from numerous uncertainties (see the text), and hence equation (7) should be used with caution. Finally, applying equation (7) to equation (2), we inferred that at redshifts $> 15$, the radiative efficiency is significantly $< 0.1$ its canonical value or that $\lambda > 1$; which suggests that SMBH's may have experienced super-Eddington accretion for a short period of $\sim$150 Myr from the inception of their seeds at $z \sim 30$ or that the radiative efficiency was $< 0.1$.

## ACKNOWLEDGEMENTS

I thank Manuel Chirouze for solving the equations using the SANN method, drafting the figures, and suggesting using density contours in Figs 1 and 2. Fabio Pacucci of the Center of Astrophysics at Harvard critically reviewed an advance copy of this article. I thank him for his review and helpful suggestions. I thank the anonymous reviewer for many helpful suggestions.

## CONFLICT OF INTEREST

The authors declare no conflict of interests.

## DATA AVAILABILITY

No new data were generated in this study.

## REFERENCES

Aggarwal Y., 2024, MNRAS, 530, 1512
Barkana R., Loeb A., 2001, Phys. Rep., 349, 125
Begelman M. C., Volonteri M., Rees M. J., 2006, MNRAS, 370, 289
Belisle C. J. P., 1992, J Appl. Probab., 29, 885
Bergström L., Goober I., 2006, Cosmology and Particle Astrophysics, Springer, Chichester, U.K. p. 77
Bhatt M. et al., 2024, A&A, 686, A141
Bhowmick A. K. et al., 2024, MNRAS, 531, 4311
Bromm V., Loeb A., 2003, ApJ, 596, 34
Chi-Hong L., Ke-Jung C., Chong-Yuan H., 2023 ApJ., 952, 121
Couchman H. M. P., Rees M. J. 1986, MNRAS, 221, 53
Davies M. B., Miller M. C., Bellovary J. M., 2011, ApJ, 740, L42
Devecchi B., Volonteri M., Rossi E. M., Colpi M., Portegies Zwart S., 2012, MNRAS, 421, 1465
Fan X., Bañados E., Simcoe R. A., 2023, ARA&A, 61, 373
Fragione G., Pacucci F., 2023, ApJ, 958, l24
Freitag M., Gäurkan M. A., Rasio F. A., 2006, MNRAS, 368, 141
Giersz M. et al., 2015, MNRAS, 454, 3150
Gravity Collaboration, 2019, A&A, 625, L10
Grazian A. et al., 2018, A&A, 613, A44
Habouzit M. et al., 2022, MNRAS, 511, 3751
Hirano S. et al., 2014, ApJ, 781, 60
Hosokawa T. et al., 2016 ApJ, 824, 119
Huang K. W., Ni Y., Feng Y., MatteoT. D.i, 2020, MNRAS, 496, 1
Johnson J. L., Bromm V., 2007, MNRAS, 374, 1557
King A. R., 2012, MNRAS, 421, 3443
Kozłowski S., 2017, ApJS, 228, 9
Larson L. R. et al., 2023, ApJ, 953, L29
Latif M. A., Ferrara A. 2016, Publ. Astron. Soc. Aust., 33, e051
Lodato G., Natarajan P., 2006, MNRAS, 371, 1813
Lupi A., Colpi M., Devecchi B., Galanti G., Volonteri M., 2014, MNRAS, 442, 3616
Madau P., Rees M. J., 2001,. ApJ, 551, L27–L30
Maiolino R. et al. 2024, Nature, 627, 59–63
Mapelli M. 2016, MNRAS,. 459, 3432
Natarajan P. et al., 2024 ApJ, 960, L1
Pacucci F., Ferrara A., Volonteri M., Dubus G., 2015,. MNRAS, 454, 3771
Pacucci F., Loeb A., 2020, ApJ, 895, 95
Planck Collaboration VI, 2020, A&A, 641, A6
Portegies Z. et al., 2004, Nature, 428, 724–726
Quataert E., Narayan R., Reid M. J. 1999, ApJ, 517, L10.
Runge J., Walker S. A., 2021, MNRAS, 502, 5487
Russell H. R., Fabian A. C., McNamara B. R., Broderick A. E., 2015, MNRAS, 451, 588
Schneider R. et al. 2023, MNRAS, 526, 3250
Shang C., Bryan G. L., Haiman Z., 2010, MNRAS, 402, 1249
Shankar F., Crocce M., Miralda-Escudé J., Fosalba P., Weinberg D. H., 2010, ApJ, 718, 231
Shen Y. et al., 2019, ApJ, 873, 35
Stacy A., Bromm V., Lee A. T., 2016, MNRAS, 462, 1307
Volonteri M., Habouzit M., Colpi M., 2021, Nat. Rev. Phys., 3, 732
Volonteri M., Rees M. J., 2005 ApJ, 633, 624
Wong K. W. et al., 2011, ApJ, 736, L23
Wong K. W. et al. 2014, ApJ, 780, 9
Wu X. B. et al., 2015, Nature, 518, 512
Wyithe J. S. B., Loeb A., 2012, MNRAS, 425, 28
Xue G.e, Zhao B.-X., Bian W.-H., Frederick G. R., 2019, AJ, 157, 14
Yuan F., Narayan R. 2014, ARA&A, 52, 529
Zubovas K., King A., 2021, MNRAS, 501, 4289

## APPENDIX

Table A1 gives the data for 59 of the 91 SMBHs in Figs 1 and 2 with references for the data sources. The BHs are listed in order of their redshift $z$ from the highest to the lowest. The first reference # is for BH's discovery paper, and the second # is for BH's mass estimate.

The remaining 32 high-$z$ SMBHs are listed at the end of the references for Table A1 in the order they appear in Table 3 of Shen et al. (2019).

J0002+2550; J0008–0626; J0810+5105; J0835+3217; J0836+0054; J0840+5624; J0841+2905; J0842+1218; J0850+3246; J1044–0125; J1137+3549; J1143+3808; J1148+5251; J1207+0630; J1243+2529; J1250+3130; J1257+6349; J1403+0902; J1425+3254; J1427+3312; J1436+5007; J1545+6028; J1602+4228; 1609+3041; J1621+5155; J1623+3112; J1630+4012; P000+26; P060+24; P210+27; P228+21; and P333+26.

**Table A1.** Parameters of SMBHS at $z > 5.6$.

| BH # | BH name | BH mass MBH ($M_{\odot}$) ($\pm 1\sigma$) | $z$ | Age (Myr) | Reference |
|---|---|---|---|---|---|
| 1 | J0313–1806 | $1.6 \times 10^{9}$ (+0.4/–0.4) | 7.64 | 676 | 1 |
| 2 | ULAS J1342+0928 | $9.1 \times 10^{8}$ (+1.3/–1.4) | 7.541 | 688 | 2 |
| 3 | J100758.264+211529.207 | $1.5 \times 10^{9}$ (+0.2/–0.2) | 7.52 | 690 | 3 |
| 4 | ULAS J1120+0641 | $2.0 \times 10^{9}$ (+1.5/–0.7) | 7.085 | 747 | 4 |
| 5 | J124353.93+010038.5 | $3.3 \times 10^{8}$ (+2.0/–2.0) | 7.07 | 749 | 5 |
| 6 | J0038–1527 | $1.33 \times 10^{9}$ (+0.25/–0.25) | 7.021 | 756 | 6 |
| 7 | DES J025216.64–050331.8 | $1.39 \times 10^{9}$ (+0.16/–0.16) | 7 | 759 | 7 |
| 8 | ULAS J2348–3054 | $2.1 \times 10^{9}$ (+0.5/–0.5) | 6.886 | 775 | 8 |
| 9 | VDES J0020–3653 | $1.67 \times 10^{9}$ (0.32/–0.32) | 6.834 | 783 | 9 |
| 10 | PSO J172.3556+18.7734 | $2.9 \times 10^{8}$ (+0.7/–0.6) | 6.823 | 784 | 10 |
| 11 | ULAS J0109–3047 | $1.5 \times 10^{9}$ (+0.4/–0.4) | 6.745 | 796 | 8 |
| 12 | HSC J1205–0000 | $2.9 \times 10^{9}$ (+0.4/–0.4) | 6.73 | 799 | 11, 12 |
| 13 | VDES J0244–5008 | $1.15 \times 10^{9}$ (+0.39/–0.39) | 6.724 | 800 | 9 |
| 14 | PSO J338.2298 | $3.7 \times 10^{9}$ (+1.3/–1.0) | 6.658 | 810 | 13 |
| 15 | ULAS J0305–3150 | $1.0 \times 10^{9}$ (+0.1/–0.1) | 6.604 | 819 | 8 |
| 16 | PSO J323.1382 | $1.39 \times 10^{9}$ (+0.32/–0.51) | 6.592 | 821 | 14 |
| 17 | PSO J231.6575 | $3.05 \times 10^{9}$ (+0.44/-2.24) | 6.587 | 820 | 14 |
| 18 | PSO J036.5078 | $3 \times 10^{9}$ (+0.92/–0.77) | 6.527 | 831 | 13, 14 |
| 19 | VDES J0224–4711 | $2.12 \times 10^{9}$ (+0.42/–0.42) | 6.526 | 831 | 9 |
| 20 | PSO J167.6415 | $3 \times 10^{8}$ (+0.08/–0.12) | 6.508 | 834 | 13, 14 |
| 21 | PSO J261+19 | $6.7 \times 10^{8}$ (+0.21/–0.21) | 6.483 | 839 | 15 |
| 22 | PSO J247.2970 | $5.2 \times 10^{8}$ (+0.22/–0.25) | 6.476 | 840 | 14 |
| 23 | PSO J011+09 | $1.20 \times 10^{9}$ (+0.51/–0.51) | 6.458 | 843 | 15 |
| 24 | CFHQS J0210–0456 | $8 \times 10^{7}$ (+5.5/–4.0) | 6.438 | 846 | 16 |
| 25 | CFHQS J2329–0301 | $2.5 \times 10^{9}$ (+0.4/–0.4) | 6.417 | 850 | 16 |
| 26 | SDSS J1148+5251 | $2.7 \times 10^{9}$ (+0.4/–0.4) | 6.41 | 851 | 17, 18 |
| 27 | HSC J0859+0022 | $3.8 \times 10^{7}$ (+0.1/–0.18) | 6.388 | 855 | 11, 19 |
| 28 | HSC J1152+0055 | $6.3 \times 10^{8}$ (+0.8/-1.2) | 6.36 | 860 | 11, 19 |
| 29 | SDSS J1148+0702 | $1.26 \times 10^{9}$ (+0.14/–0.14) | 6.339 | 863 | 20 |
| 30 | SDSS J1030+0524 | $1.0 \times 10^{9}$ (+0.2/–0.2) | 6.3 | 870 | 21, 22 |
| 31 | SDSS J0100+2802 | $1.24 \times 10^{10}$ (+0.19/–0.19) | 6.3 | 870 | 23 |
| 32 | CFHQS J0050+3445 | $2.6 \times 10^{9}$ (+0.50/–0.4) | 6.253 | 879 | 16 |
| 33 | HSC J2239+0207 | $1.1 \times 10^{9}$ (+3/-2) | 6.245 | 880 | 19 |
| 34 | VDES J0330–4025 | $5.87 \times 10^{9}$ (+0.89/–0.89) | 6.239 | 881 | 15 |
| 35 | VDES J0323–4701 | $5.5 \times 10^{8}$ (+1.26/-1.26) | 6.238 | 881 | 15 |
| 36 | SDSS J1623+3112 | $1.5 \times 10^{9}$ (+0.3/–0.3) | 6.211 | 886 | 21 |
| 37 | SDSS J1048+4637 | $3.9 \times 10^{9}$ (+2.1/-2.1) | 6.198 | 889 | 24 |
| 38 | PSO J359–06 | $1.66 \times 10^{9}$ (+0.21/–0.21) | 6.164 | 895 | 15 |
| 39 | CFHQS J0221–0802 | $7 \times 10^{8}$ (+7.5/-4.7) | 6.161 | 896 | 16 |
| 40 | HSC J1208–0200 | $7.1 \times 10^{8}$ (+2.4/-5.2) | 6.144 | 899 | 19 |
| 41 | ULAS J1319+0950 | $2.7 \times 10^{9}$ (+0.6/–0.6) | 6.13 | 902 | 25, 26 |
| 42 | CFHQS J1509–1749 | $3 \times 10^{9}$ (+0.3/–0.3) | 6.121 | 903 | 16 |
| 43 | PSO J239–07 | $3.63 \times 10^{9}$ (+0.20/–0.20) | 6.114 | 905 | 15 |
| 44 | HSC J2216–0016 | $7 \times 10^{8}$ (+1.4/-2.3) | 6.109 | 906 | 19 |
| 45 | CFHQS J2100–1715 | $3.37 \times 10^{9}$ (+0.64/–0.64) | 6.087 | 910 | 16 |
| 46 | SDSS J0303–0019 | $3 \times 10^{8}$ (+2.0/–2.0) | 6.079 | 911 | 24 |
| 47 | SDSS J0353+0104 | $1.4 \times 10^{9}$ (+1.0/–1.0) | 6.072 | 913 | 24 |
| 48 | SDSS J0842+1218 | $1.7 \times 10^{9}$ (+1.2/–1.2) | 6.069 | 913 | 24 |
| 49 | SDSS J1630+4012 | $9 \times 10^{8}$ (+0.8/–0.8) | 6.058 | 915 | 24 |
| 50 | PSO J158–14 | $2.15 \times 10^{9}$ (+0.25/–0.25) | 6.057 | 916 | 15 |
| 51 | CFHQS J1641+3755 | $2.4 \times 10^{8}$ (+1.0/–0.8) | 6.047 | 918 | 16 |
| 52 | SDSS J1306+0356 | $1.1 \times 10^{9}$ (+0.1/–0.1) | 6.017 | 923 | 21 |
| 53 | SDSS J2310+1855 | $2.8 \times 10^{9}$ (+0.6/–0.6) | 6.003 | 926 | 19, 27 |
| 54 | CFHQS J0055+0146 | $2.4 \times 10^{8}$ (+0.9/–0.7) | 5.983 | 930 | 16 |
| 55 | PSO J056–16 | $7.5 \times 10^{8}$ (+0.07/–0.07) | 5.975 | 932 | 15 |

**Table A1** – *continued*

| BH # | BH name | BH mass MBH ($M_{\odot}$) ($\pm 1\sigma$) | $z$ | Age (Myr) | Reference |
|---|---|---|---|---|---|
| **56** | **SDSS J1411+1217** | **1.1 × $10^9$ (+0.1/–0.1)** | **5.93** | **941** | **22, 28** |
| **57** | **SDSS J0005–0006** | **3 × $10^8$ (+0.1/–0.1)** | **5.85** | **957** | **22, 28** |
| **58** | **SDSS J0836+0054** | **2.7 × $10^9$ (+0.6/–0.6)** | **5.82** | **964** | **22, 28** |
| **59** | **SDSS J1044–0125** | **1.05 × $10^{10}$ (+0.16/–0.16)** | **5.784** | **971** | **21, 29** |

*Notes*. References in Table A1 are follows: [1] Wang, F., et al., 2021, ApJL 907, L1. [2] Bañados, E. et al. Nature, 2017, 553, 473. [3] Yang, J., et al. 2020, ApJL. 897, L14. [4] Mortlock, D.J., et al. 2011, Nature, 474, 616. [5] Matsuoka, Y., et al. 2019, ApJL 872, L2. [6] Wang F., et al. 2018, ApJL 869, L9. [7] Wang, F., et al. 2020, ApJ. 896, 23. [8] Venemans, B.P., et al. 2013, ApJ. 779, 24. [9] Reed, S.L., et al. 2019, MNRAS. 2, 1874–1885. [10] Bañados, E. et al. 2021, ApJ. 909, 80. [11] Matsuoka, Y., et al. 2016, ApJ. 828, 26. [12] Kato, N., et al. 2020, ASJ. 72, 5, 84. [13] Venemans B.P., et al. 2015, ApJL 801, L11. [14] Mazzucchelli, C., et al. 2017, 849, 91. [15] Eilers, A.C., et al. 2020, ApJ. 900, 37. [16] Willott, C.J., et al. 2010, ApJ.140, 546. [17] Willott, C.J., McLure, R.J., Jarvis, M.J., 2003, ApJL 587, L15. [18] Gallerani S. et al. 2017, MNRAS. 467, 3590. [19] Onoue, M., et al. 2019, ApJ. 880, 77. [20] Jiang, L., et al. 2016, ApJ. 833, 222. [21] Jiang, L., et al. 2007, ApJ. 134, 1150. [22] Kurk, J.D., et al. 2007, ApJ. 669, 32. [23] Wu, X.B., et al. 2015, Nature, 518, 512. [24] De Rosa G. et al. 2011, ApJ. 739, 56. [25] Mortlock, D.J., et al, 2009, A&A, 505, 97. [26] Y. Shao, Y., et al. 2017, ApJ. 845, 138. [27] Wang, F., et al. 2011, ApJL. 739, L34. [28] Fan, X., et al. 2006, ApJ. 131, 1203. [29] Fan, X., et al. 2000, ApJ. 120, 1167.